\documentclass[aps,prl,reprint,superscriptaddress]{revtex4-1}
\usepackage{amsmath,amssymb,amsthm}
\usepackage{braket}
\usepackage{color}
\usepackage{graphicx}
\usepackage{bm}
\usepackage{comment}
\graphicspath{}
\begin{document}
	\title{Variational Quantum Gate Optimization}
	
	\author{Kentaro Heya}
	\email[]{kheya@qc.rcast.u-tokyo.ac.jp}
	\affiliation{Research Center for Advanced Science and Technology (RCAST), The University of Tokyo, Meguro-ku, Tokyo 153-8904, Japan}
	
	\author{Yasunari Suzuki}
	\affiliation{NTT Secure Platform Laboratories, NTT Corporation, Musashino 180-8585, Japan}
	
	\author{Yasunobu Nakamura}
	\affiliation{Research Center for Advanced Science and Technology (RCAST), The University of Tokyo, Meguro-ku, Tokyo 153-8904, Japan}
	\affiliation{Center for Emergent Matter Science (CEMS), RIKEN, Wako, Saitama 351-0198, Japan}
	
	\author{Keisuke Fujii}
	\affiliation{Graduate School of Science, Kyoto University, Sakyo-ku, Kyoto 606-8302, Japan}
	\affiliation{JST PRESTO, Kawaguchi, Saitama 332-0012, Japan}
	\date{\today}
	
	\begin{abstract}
		We propose a gate optimization method, which we call variational quantum gate optimization (VQGO).
		VQGO is a method to construct a target multi-qubit gate by optimizing a parametrized quantum circuit which consists of tunable single-qubit gates with high fidelities and fixed multi-qubit gates with limited controlabilities.
		As an example, we apply the proposed scheme to the models relevant to superconducting qubit systems.
		We show in numerical simulations that the high-fidelity CNOT gate can be constructed with VQGO using cross-resonance gates with finite crosstalk.
		We also demonstrate that fast and a high-fidelity four-qubit syndrome extraction can be implemented with simultaneous cross-resonance drives even in the presence of non-commutative crosstalk.
		VQGO gives a pathway for designing efficient gate operations for quantum computers.
	\end{abstract}
	
	\maketitle
	
	Hybrid quantum-classical (HQC) algorithms aim at realizing quantum advantage in shallow depth quantum circuits with an aid of classical computation. Recently, HQC algorithms have been extensively studied with the expectations that they may solve practical problems in the near future~\cite{preskill2018quantum,peruzzo2014variational,kandala2017hardware,colless2018computation,farhi2014quantum,otterbach2017unsupervised,mitarai2018quantum,vojtech2018quantum,wecker2015solving,mcclean2016theory,wang2017experimental,guerreschi2017practical,mcclean2017hybrid,romero2018strategies,johnson2017qvector,khatri2018quantum}.
	For example, calculation of the ground-state energy of a given Hamiltonian has been performed with an algorithm called variational quantum eignsolver (VQE)~\cite{peruzzo2014variational,kandala2017hardware,colless2018computation}.
	Also, supervised machine learning applicable to problems such as classification has been performed with an algorithm called quantum circuit learning (QCL)~\cite{mitarai2018quantum,vojtech2018quantum}.
	
	Quantum gate fidelities directly limit the sizes of executable problems in quantum computers without quantum error correction.
	While HQC algorithms require less quantum gates, the state-of-the-art fidelities of quantum gates are still insufficient to deal with practical problems. 
	In superconducting qubit systems, for example, single-qubit gate fidelities reach the level of $0.999$. However, fidelities of two-qubit gates are still around $0.95$--$0.99$~\cite{takita2017experimental,kelly2015state,corcoles2015demonstration,barends2014logic,takita2016demonstration,mckay2017efficient}. 
	Quantum error correction can, in principle, improve these fidelities.
	However, that also requires high physical gate fidelities far beyond the threshold value~\cite{bravyi1998quantum,fowler2012surface}.
	Thus, a method to improve the physical gate fidelities is strongly demanded.
	
	In this Letter, we propose a scheme to exploit a HQC algorithm itself to improve gate fidelities by optimizing multi-qubit gate operations using near-term quantum devices.
	In near-term quantum devices, gate fidelities are finally limited by the coherent time.
	Therefore, for implementing a given unitary operation on a set of qubits, it is strongly demanded to find an efficient gate construction.
	So far, variational gate optimization methods have been mostly studied in the context of quantum optimal control (QOC)~\cite{werschnik2007quantum}, where gate fidelities are improved with optimally shaped control pulses derived from numerical simulations on a classical computer or actual measurements on the quantum devices under test.
	A variety of QOC methods have been proposed theoretically~\cite{PhysRevLett.106.190501,PhysRevA.92.062343,khaneja2005optimal,li2017hybrid,konnov1999global,sklarz2002loading,reich2012monotonically,machnes2018tunable,kirchhoff2018optimized,sorensen2018gradient,Nocedal2006NO,tannor1985control,conolly1986optimal,peirce1988optimal,zahedinejad2015high,wu2018data} and used in experiments~\cite{kosloff1989wavepacket,brixner2001photoselective,bartels2000shaped,buckup2006singlet,haessler2014optimization,dolde2014high,heeres2017implementing,li2017hybrid,feng2018closed,lu2017enhancing}.
	In those methods, the pulse shapes are approximated with some ansatz functions.
	While the ansatz functions should have high representation power in order to faithfully generate the target gates, a large number of parameters to label them make the optimization difficult. Therefore, pulse decomposition methods with a reduced number of variables have been developed as compromised solutions.
	
	Here, we propose a HQC algorithm for gate optimization, variational quantum gate optimization (VQGO).
	The key idea of VQGO is to iteratively optimize quantum gates in the form of parametrized quantum circuits.
	VQGO is a variational method to construct a given target multi-qubit gate from source gates, which are driven by natural interaction in the system Hamiltonian under a drive, and single-qubit gates.
	Only the single-qubit gates are parametrized with the rotation angles, which are thought to be high fidelity.
	The target multi-qubit gate is constructed by iteratively optimizing the rotation angles based on measured values of a cost function.
	We perform detailed numerical simulation on two models related to superconducting qubit systems.
	We show that even in the presence of crosstalk VQGO can achieve higher fidelity of the CNOT gate compared with a conventional method~\cite{rigetti2010fully}.
	We also implement a fast and high-fidelity four-qubit syndrome extraction, which is often required in quantum error correction, with simultaneous cross-resonance drives in the presence of non-commutative crosstalk.
	
\begin{figure}[t]
		\begin{center}
			\includegraphics[width=8 cm]{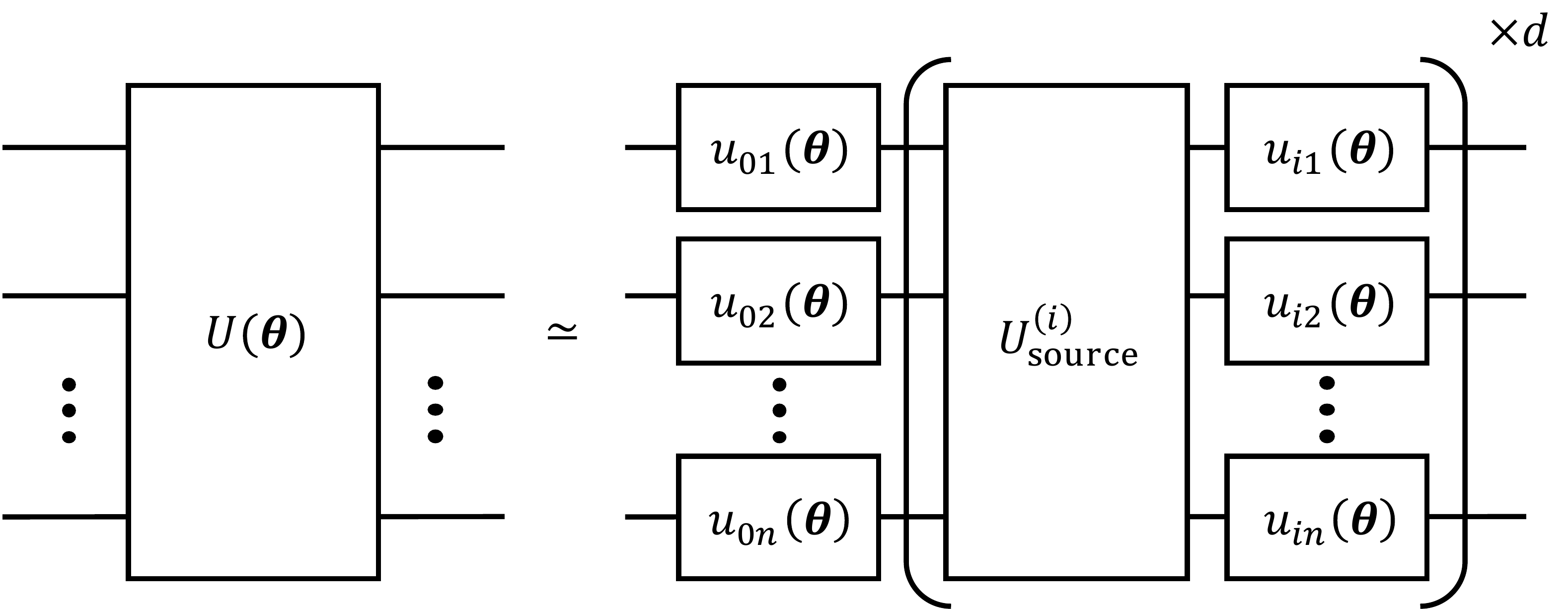}
			\caption{Parametrized quantum circuit in VQGO. Source gates $U^{(i)}_{\rm{source}}$ and tensor products of single qubit gates $\bigotimes_{j=1}^n u _{ij}(\bm{\theta})$ are alternatively repeated $d$ times.}
			\label{gate_const}
		\end{center}
	\end{figure}

	{\it Variational Quantum Gate Optimization.}
	VQGO is a method of gate optimization to construct an $n$-qubit target gate $U_{\rm{target}}$ with a parametrized quantum circuit which consists of single-qubit gates and source gates.
	The protocol of VQGO is as follows.
	We first choose $d$ source gates $\{U_{\rm source}^{(i)}\}_{i=1}^d$, where $d$ is the depth of the parametrized quantum circuit.
	We use $n(d+1)$ single-qubit gates parametrized with $3n(d+1)$ parameters $\bm{\theta}\equiv\{\theta_{ijk}\}$, $\theta_{ijk}\in [0,2\pi)$, $0\le i \le d$, $1\le j \le n$, and $k \in \{0,1,2\}$.
	Each single-qubit gate is written as
	\begin{align}
	u_{ij}(\bm{\theta})=e^{-i\theta_{ij0} \sigma_x}e^{-i\theta_{ij1} \sigma_y}e^{-i\theta_{ij2} \sigma_x}.
	\end{align}
	Note that this formalism can represent an arbitrary single-qubit gate.
	Then, the parametrized quantum circuit in VQGO is defined as 
	\begin{align}
	U(\bm{\theta})=&\left(\bigotimes_{j=1}^n u_{0j}(\bm{\theta})\right)U^{(1)}_{\rm{source}}\left(\bigotimes_{j=1}^n u_{1j}(\bm{\theta})\right)\cdots\nonumber\\
	&\cdots \left(\bigotimes_{j=1}^n u_{(d-1)j}(\bm{\theta})\right)U^{(d)}_{\rm{source}}\left(\bigotimes_{j=1}^n u_{dj}(\bm{\theta})\right),
	\end{align}
	which is sketched in Fig.\,\ref{gate_const}. 
	
\begin{figure}[t]
		\begin{center}
			\includegraphics[width=6 cm]{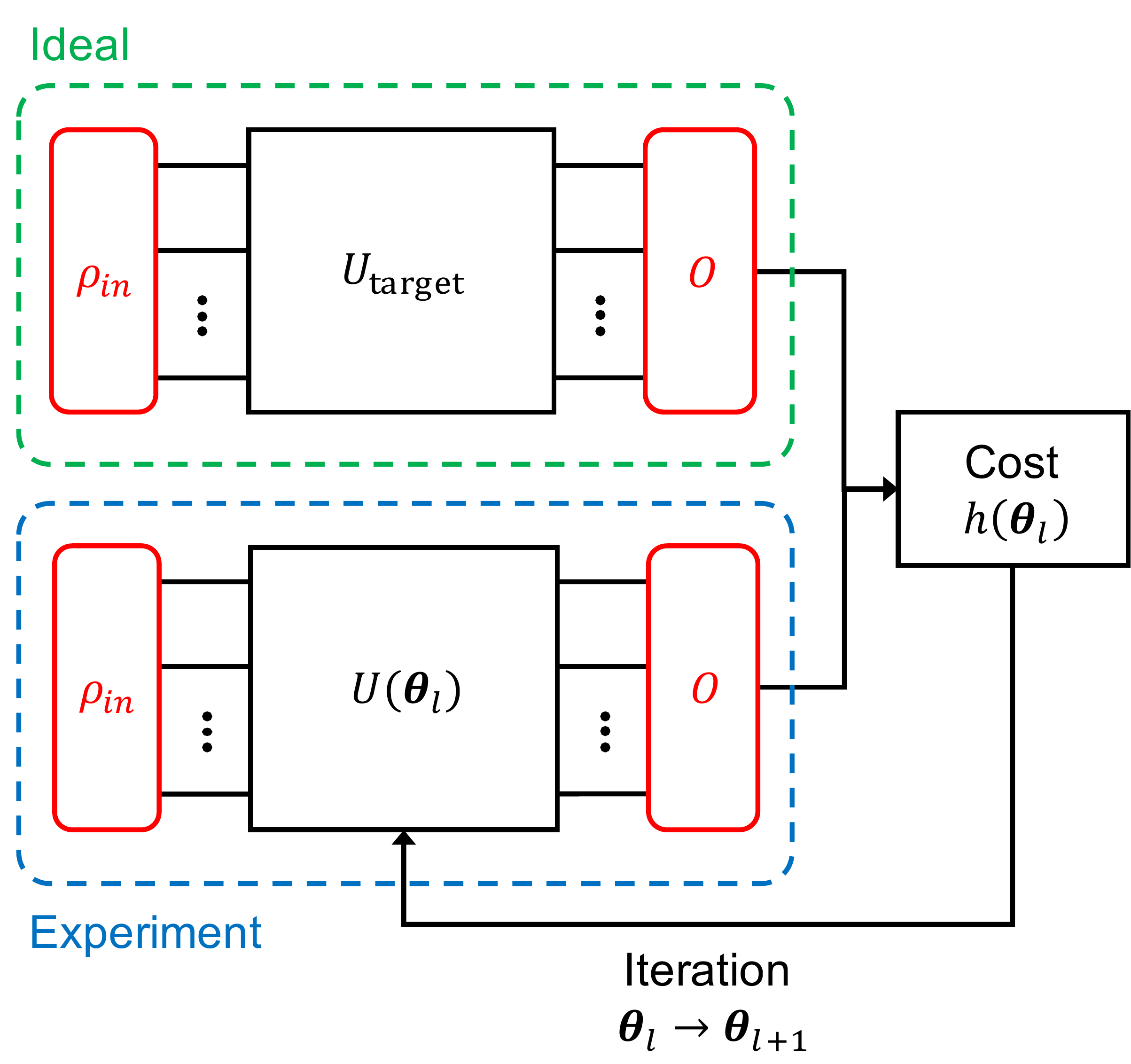}
			\caption{Iterative optimization protocol of VQGO. A cost function such as averaged gate infidelity is calculated for the $l$-th parametrized quantum circuit $U\left(\bm{\theta}_l\right)$ from the tomography measurement based on a complete set of input states $\bm{\rho}_{\rm in}$ and measurement bases $\bm{O}$. Then, parameters $\bm{\theta}_l$ are updated to $\bm{\theta}_{l+1}$.}
			\label{circuit_const}
		\end{center}
	\end{figure}

	We approximate the target gate $U_{\rm target}$ using the parameterized quantum circuit $U(\bm{\theta})$ by optimizing the parameters $\bm{\theta}$ iteratively.
	For the optimization, we choose a cost function $h(\bm{\theta})$ which evaluates a distance between the quantum channels generated by the unitary operators $U(\bm{\theta})$ and $U_{\rm{target}}$ in terms of the given parameters $\bm{\theta}$. 
	The optimization is started with the initial parameters $\bm{\theta}_{l = 0}$, each element of which is uniformly sampled from $[0,2\pi)$.
	In each iteration, the parameters are updated from ${\bm \theta}_l$ to ${\bm \theta}_{l+1}$ so as to minimize the cost $h\left({\bm \theta}\right)$ with gradient-free or gradient-based optimizers as shown in Fig.\,\ref{circuit_const}.
	When we use a gradient-based optimizer, we assume the cost function is a function of the expectation values $\bm{E}(\bm{\theta})=\left(E_1(\bm{\theta}),E_2(\bm{\theta})\cdots\right)$, where $E_m(\bm{\theta})\equiv\mathrm{Tr}[O^{(m)} U(\bm{\theta}) \rho^{(m)}_{\rm in} U^{\dagger} (\bm {\theta})]$. The observables $O^{(m)}$ and density operators $\rho^{(m)}_{\rm in}$ are chosen depending on the cost function $h$. Then, we can estimate the gradient of the cost, $\nabla_{\theta_{ijk}} h(\bm{\theta})$ with respect to $\theta_{ijk} \in {\bm \theta}$ as
	\begin{align}
	\frac{\partial}{\partial \theta_{ijk}} h(\bm{\theta})=\frac{\bm{E}\left(\bm{\theta}_{ijk}^{+}\right)-\bm{E}\left(\bm{\theta}_{ijk}^{-}\right)}{2}\cdot\nabla_{\bm{E}(\bm{\theta})}h(\bm{\theta}),
	\end{align}
	where
	\begin{align}
	\bm{\theta}_{ijk}^{\pm}&\equiv\{\theta^{\pm}_{i'j'k'}\},\\
	\theta^{\pm}_{i'j'k'}&=\theta_{ijk}\pm\frac{\pi}{2}\delta_{i,i'}\delta_{j,j'}\delta_{k,k'},
	\end{align}
	with $0\le i' \le d$, $1\le j' \le n$, $k' \in \{0,1,2\}$, and Kronecker delta $\delta_{a,a'}$~($a={i,j,k}$)~\cite{mitarai2018quantum,khaneja2005optimal}.
	We iteratively repeat this update until the cost function converges.
	
	We adopt average gate infidelity (AGI) as the cost function in the optimization. AGI represents how far in average the two quantum channels are for pure states $\psi$ randomly sampled according to the uniform Haar measure~\cite{nielsen2002quantum}. Then, the cost function is represented as 
	\begin{align}
	h\left( {\bm \theta} \right)
	&= 1-\int \mathrm{Tr}\left[\mathcal{U}_{\rm{target}}\left(\psi\right)\mathcal{U}_{\bm{\theta}}\left(\psi\right)\right] d\mu_\psi,
	\end{align}
	where $\mathcal{U}_{\rm{target}}$ and $\mathcal{U}_{\bm{\theta}}$ are superoperators corresponding to $U_{\rm{target}}$ and $U(\bm{\theta})$, respectively. 
	Note that the AGI is evaluated by direct fidelity estimation~\cite{PhysRevLett.106.230501,heya2018supli}, which calculates the AGI from the measurement outcomes more efficiently than full process tomography.
	
	There are two essential factors which determine the performance of VQGO.
	The first is the choice of the source gates and the depth $d$.
	Even when the cost value converges, the target gates are not necessarily well approximated. For example, a parametrized quantum circuit composed of source gates without entangling power~\cite{zanardi2001entanglement} such as SWAP gates cannot approximate any entangling target gate. On the other hand, it is known that a parametrized quantum circuit with CNOT gates, which have maximum entangling power, as the source gates can approximate an arbitrary unitary operator in SU$(4)$ at depth $d=3$~\cite{sousa2006universal}.
	For SU$(4)$, a detailed discussion of the relation between the representation power of parametrized quantum circuits and their source gates is given in Supplementary Information~\cite{heya2018supli}.
	
	The other factor is the choice of the optimizer. There are a number of possible choices of gradient-free~\cite{nelder1965simplex,powell1964efficient,conn1997convergence} and gradient-based~\cite{fletcher2013practical,byrd1995limited,hestenes1952methods,kingma2014adam,119632} optimizers. We tried a variety of existing optimizers and found that a gradient-based optimizer L-BFGS-B~\cite{byrd1995limited} shows the fastest convergence to an optimal value, even if we take into account the number of measurements necessary for the gradient evaluation. It is known that there exist only a few local traps in quantum gate optimization in general~\cite{russell2018control}. Thus, we expect local traps do not largely affect the performance of VQGO.
	
	{\it Examples of VQGO.}
	Below we consider two examples of VQGO applied to the problems in superconducting qubit systems. One is the gate optimization of the CNOT gate in a two-qubit system, and the other is that of a four-qubit syndrome extraction in a five-qubit system. Both of them are implemented with mutually detuned qubits connected in a nearest-neighbor way.
	
	{\it CNOT gate in two-qubit system.}
	First, we show VQGO can be used for generating the CNOT gate from other possibly imperfect two-qubit gates with undesired interaction terms. We consider two coupled superconducting qubits detuned from each other and labeled $\rm{Q_1}$ and $\rm{Q_2}$, respectively. The entangling source gates are provided by driving $\rm{Q_1}$ with the eigenfrequency of $\rm{Q_2}$, which are called the cross-resonance (CR) gates~\cite{rigetti2010fully,chow2011simple}. A simplified drive Hamiltonian of the CR gate under the rotating wave approximation is written as
	\begin{align}
	\mathcal{H}\left(\Omega\right)=&\delta\hat{\sigma}^+_1 \hat{\sigma}^-_1 + g\left(\hat{\sigma}^+_1 \hat{\sigma}^-_2 + \hat{\sigma}^-_1 \hat{\sigma}^+_2 \right)\nonumber\\
	& + \frac{\Omega}{2}\left[\left(\hat{\sigma}^-_1 + \hat{\sigma}^+_1 \right) + \varepsilon\left(e^{-i\phi}\hat{\sigma}^-_2 + e^{i\phi}\hat{\sigma}^+_2\right)\right],
	\end{align}
	where $\hat{\sigma}^-_i$  ($\hat{\sigma}^+_i$) is the lowering (raising) operator of $\rm{Q_i}$, $\delta$ and $g$ are the detuning and coupling strengths between the qubits, respectively, and $\Omega$ is the CR drive amplitude. The parameters $\varepsilon$ and $\phi$ are the amplitude attenuation and phase delay in crosstalk of the CR drive, respectively, which is usually caused by the leakage of the strong CR drive pulse to the neighboring qubit.
	Note that crosstalk in single-qubit gates is ignored as the effect is much weaker than that of the CR drive.
	The CR gate is generated by the time evolution under the CR Hamiltonian namely,
	\begin{align}
	U_{\rm{CR}}\left(\Omega,t\right)=e^{-{\rm{i}}\mathcal{H}\left(\Omega\right)t},
	\end{align}
	where $t$ is the gate time.
	Our purpose here is to construct the CNOT gate with the CR gates as source gates.
	
\begin{figure}[t]
		\begin{center}
			\includegraphics[width=8 cm]{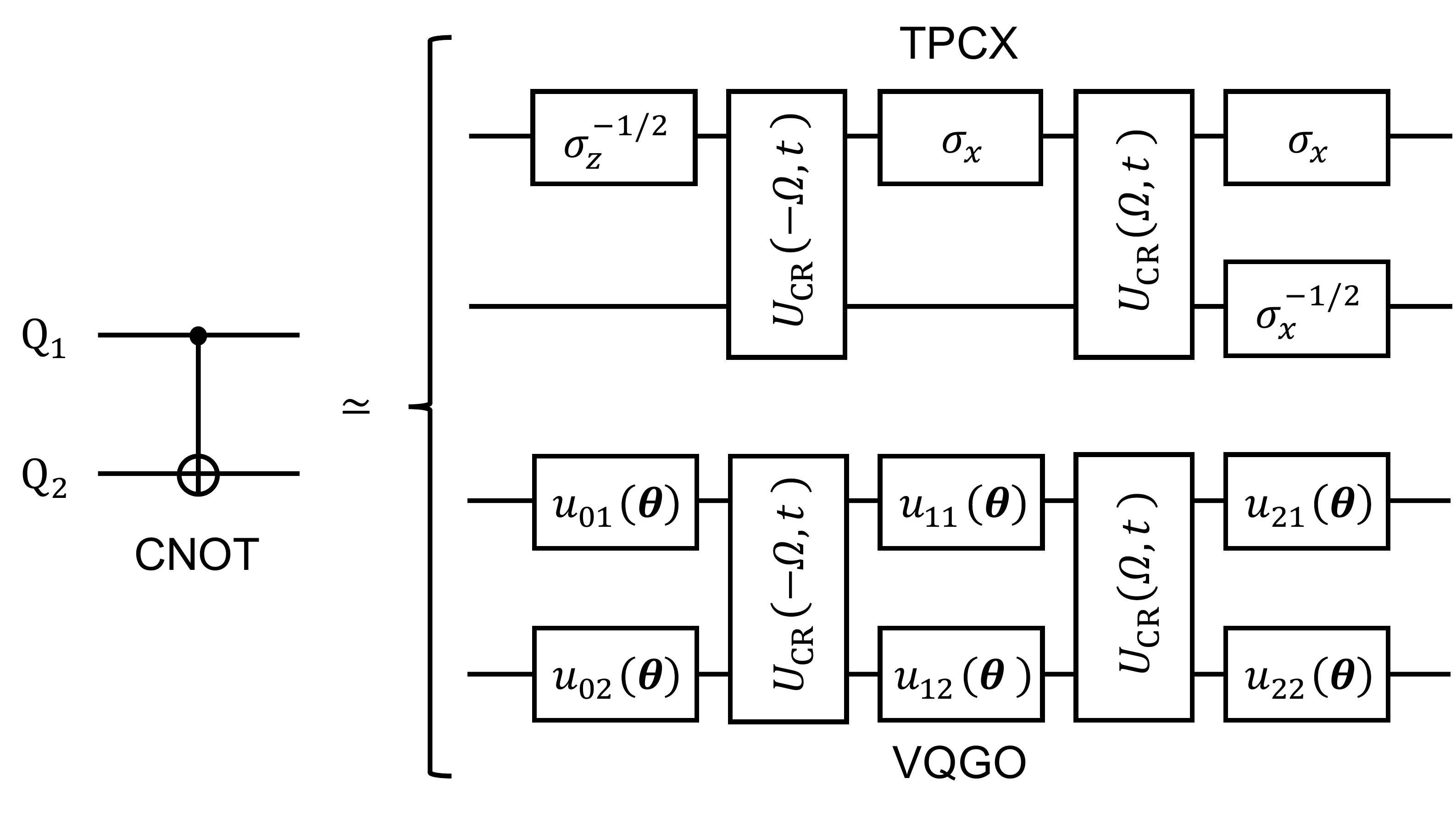}
			\caption{Quantum circuits of TPCX and VQGO. In each method, the circuit consists of two opposite-phase CR gates $U_{\rm{CR}}\left(\pm\Omega,t\right)$ and single-qubit gates. While the single-qubit gates in TPCX are given analytically, those in VQGO are parametrized and tunable.}
			\label{cnot_vqgo}
		\end{center}
	\end{figure}

	We compare VQGO with another method to generate the CNOT gate using two CR gates, i.e., the two-pulse echoed cross-resonance CNOT gate (TPCX)~\cite{takita2017experimental}.
	In TPCX, the CNOT gate is constructed with two CR gates generated by opposite-phase drives and fixed single-qubit gates.
	TPCX can be regarded as an instance of parametrized quantum circuits in VQGO with depth $d=2$.
	For a fair comparison with TPCX, we use a parametrized quantum circuit with the same set of the source gates as TPCX, i.e., two opposite-phase CR gates, in VQGO. The gate constructions in TPCX and VQGO are shown in Fig.\,\ref{cnot_vqgo}.
	
	In the numerical calculations, we choose $\delta=200\ \rm{MHz}$, $g=5\ \rm{MHz}$, and $t=75\ \rm{ns}$ as typical values in experiments~\cite{sheldon2016procedure}.
	For comparison, we calculate the cases with and without crosstalk.
	As for crosstalk, we simulate two different conditions with the amplitude attenuation $\varepsilon=0.1$ and $1.0$ for the phase delay $\phi=\pi/4$.
	Here, we also optimize the source gate drive amplitudes to obtain the best performance at $t=75\ \rm{ns}$ for TPCX and VQGO, respectively, by using a gradient-free optimization method known as COBYLA~\cite{conn1997convergence}. 
	In the case of VQGO, we perform a concatenated optimization of the parameters $\bm{\theta}$ in the parametrized quantum circuit and the drive amplitudes $\Omega$ of the soruce gates. We can obtain the optimal AGI for fixed drive amplitudes by optimizing only the parameters in the circuit. We can optimize the drive amplitudes by using this AGI as a cost.
	For the optimization, we impose a constraint so that the drive amplitude does not take unrealistically large values.
	We repeat this concatenated optimizations until it converges at the values listed in Table\,\ref{params_1}.
	\begin{table}[t]
		\caption{Set of parameters used in the numerical simulation of TPCX and VQGO for the CNOT gate.}
		\label{params_1}
		\begin{ruledtabular}
			\begin{tabular}{c|cc}
				& TPCX	& VQGO	\\ \hline
				$\Omega\ (\rm{MHz})\ (\varepsilon=0)$		& 63.5 	& 77.5 		\\
				$\Omega\ (\rm{MHz})\ (\varepsilon=0.1)$ 	& 36.4 	& 114 		\\
				$\Omega\ (\rm{MHz})\ (\varepsilon=1.0)$ 	& 69.2 	& 115 		\\
			\end{tabular}
		\end{ruledtabular}
	\end{table}
	\begin{figure}[t]
		\begin{center}
			\includegraphics[width=8cm]{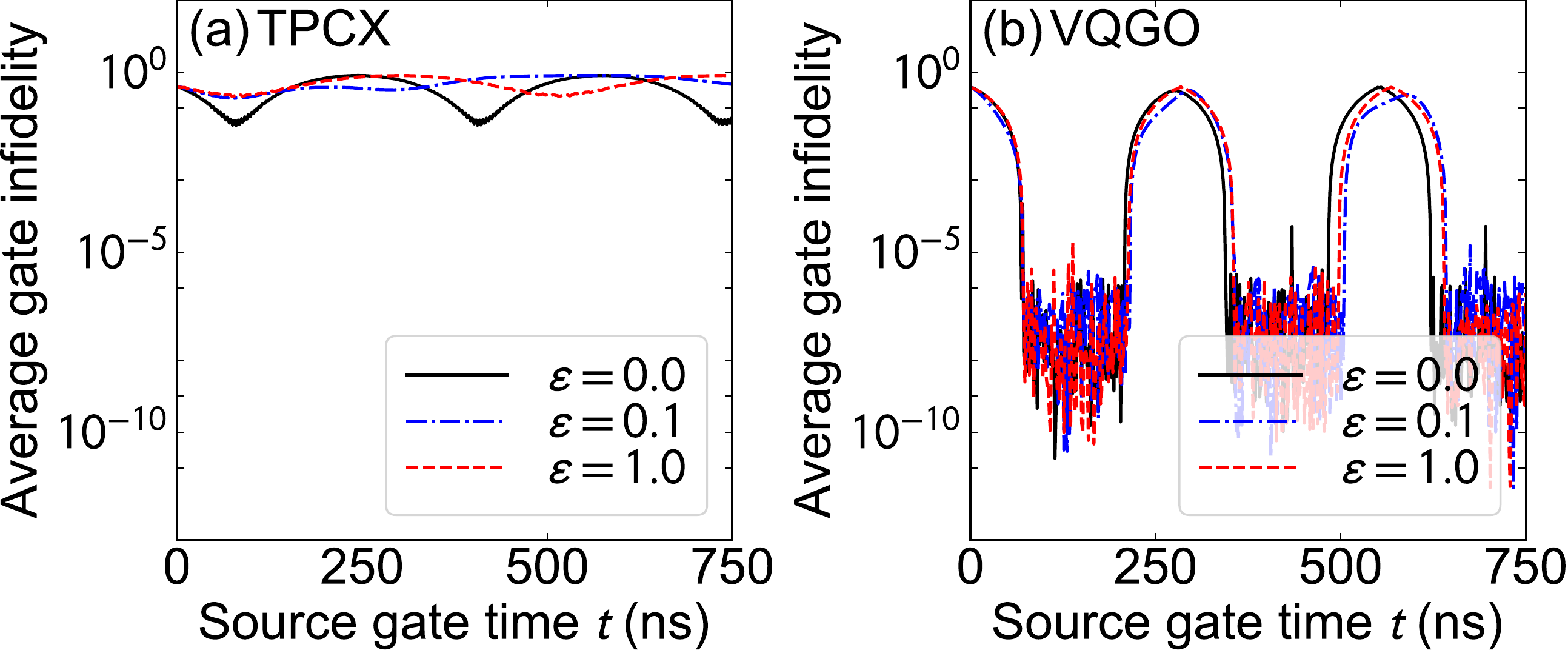}
			\caption{Average gate infidelities of generated CNOT gates as a function of the gate time $t$ of each source gate for (a)~TPCX and (b)~VQGO. The corresponding crosstalk parameters are $\varepsilon=0$ (black solid curve), $0.1$ (blue dash-dotted), and $1$ (red dashed).}
			\label{cr_result}
		\end{center}
	\end{figure}
	Under the drive amplitude $\Omega$, we sweep the gate time $t$ of the source gates from $0$ to $750\ \rm{ns}$ and plot the AGI of the obtained CNOT gate in Fig.\,\ref{cr_result}.
	For both TPCX and VQGO, the AGI shows a periodic behavior. This is due to the periodic dependence of the source gate properties on the gate time.
	In the case of TPCX, the AGI at $t=75\ \rm{ns}$ is $0.034$ even if we ignore crosstalk. If we take into account crosstalk, the AGI increases to $0.190$ for $\varepsilon=0.1$ and $0.210$ for $\varepsilon=1.0$. This is because TPCX employs approximations to construct the CNOT gate and cannot mitigate the effect of crosstalk. Note that there exists a heuristic optimization method of source gates specialized to TPCX, which can cancell crosstalk by applying an additional pulse during CR dynamics~\cite{sheldon2016procedure}.
	In contrast, even in the presence of crosstalk, the AGI of VQGO has wide regions of $t$ where the AGI converges almost to zero.
	In this region, the condition required to approximate the CNOT gate with the parametrized quantum circuit is satisfied. The details are shown in Supplementary Information~\cite{heya2018supli}.
	
\begin{figure}[t]
		\begin{center}
			\includegraphics[width=8 cm]{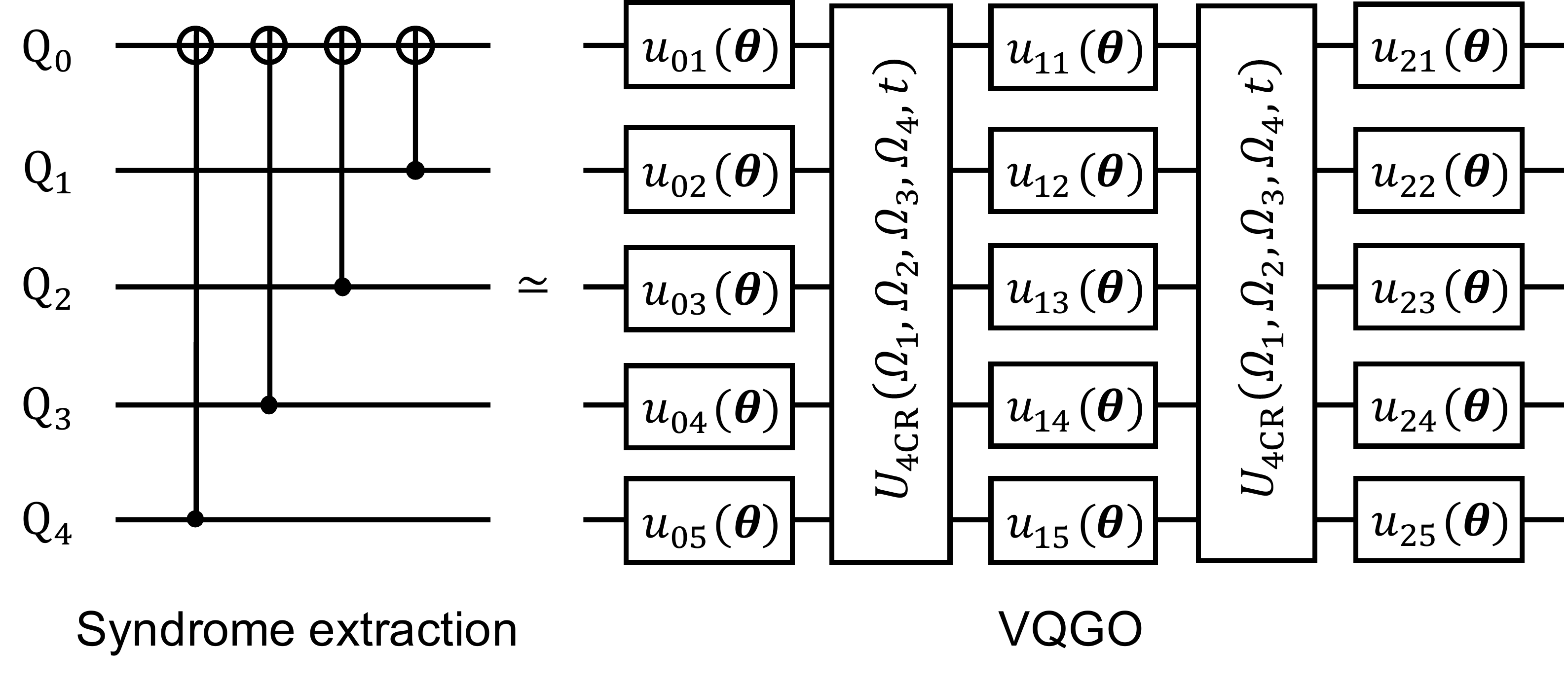}
			\caption{Parametrized quantum circuit for the four-qubit syndrome extraction.}
			\label{parity_meas}
		\end{center}
	\end{figure}

	{\it Four-qubit syndrome extraction in five-qubit system.}
	Next, we show that VQGO can be applied to a four-qubit syndrome extraction for a syndrome measurement, which is required in quantum error correction with the stabilizer codes~\cite{gottesman1997stabilizer}. In the case of the surface code on a square lattice, we need to evaluate the parity in Pauli-$Z$ basis (or $X$~basis) on the four data qubits labeled as $\rm{Q_1}$, $\rm{Q_2}$, $\rm{Q_3}$, and $\rm{Q_4}$ through a measurement of the measurement qubit labeled $\rm{Q_0}$.
	As shown in Fig.\,\ref{parity_meas}, the four-qubit syndrome extraction can be decomposed into four sequential CNOT gates with each data qubit and the measurement qubit as the control and the target, respectively.
	In conventional methods, each of the four CNOT gates is constructed with CR gates using TPCX or its variations~\cite{takita2017experimental}.
	If we perform the four CNOT gates simultaneously, it does not work appropriately because of the undesirable interactions between the parallel CR drives. In contrast, VQGO can treat such complex coupled dynamics as source gates. 
	Here, we construct the four-qubit syndrome extraction scheme with VQGO using four simultaneous CR dynamics. The simplified drive Hamiltonian under the rotating wave approximation is written as
	\begin{align}
	&\mathcal{H}\left(\Omega_1,\Omega_2,\Omega_3,\Omega_4\right)=\nonumber\\
	&\sum_{i=1}^4
	\Biggl\{
	\delta_i \hat{\sigma}^+_i \hat{\sigma}^-_i
	+ g_i \left(\hat{\sigma}^+_0 \hat{\sigma}^-_i+\hat{\sigma}^+_i \hat{\sigma}^-_0 \right)
	\nonumber\\ 
	& + 
	\frac{\Omega_i}{2}
	\left[
	\left(\hat{\sigma}^+_i + \hat{\sigma}^-_i\right) + \varepsilon_i\left(e^{-i\phi_i}\hat{\sigma}^-_0+e^{i\phi_i}\hat{\sigma}^+_0\right)
	\right]
	\Biggr\}.
	\end{align}
	The source gates are generated by the time evolution with the above Hamiltonian, namely,
	\begin{align}
	U_{\rm{4CR}}\left(\Omega_1,\Omega_2,\Omega_3,\Omega_4,t\right)=e^{-{\rm{i}}\mathcal{H}\left(\Omega_1,\Omega_2,\Omega_3,\Omega_4\right)t},
	\end{align}
	where $t$ is the gate time of each source gate.
	
\begin{figure}[t]
		\begin{center}
			\includegraphics[width=4 cm]{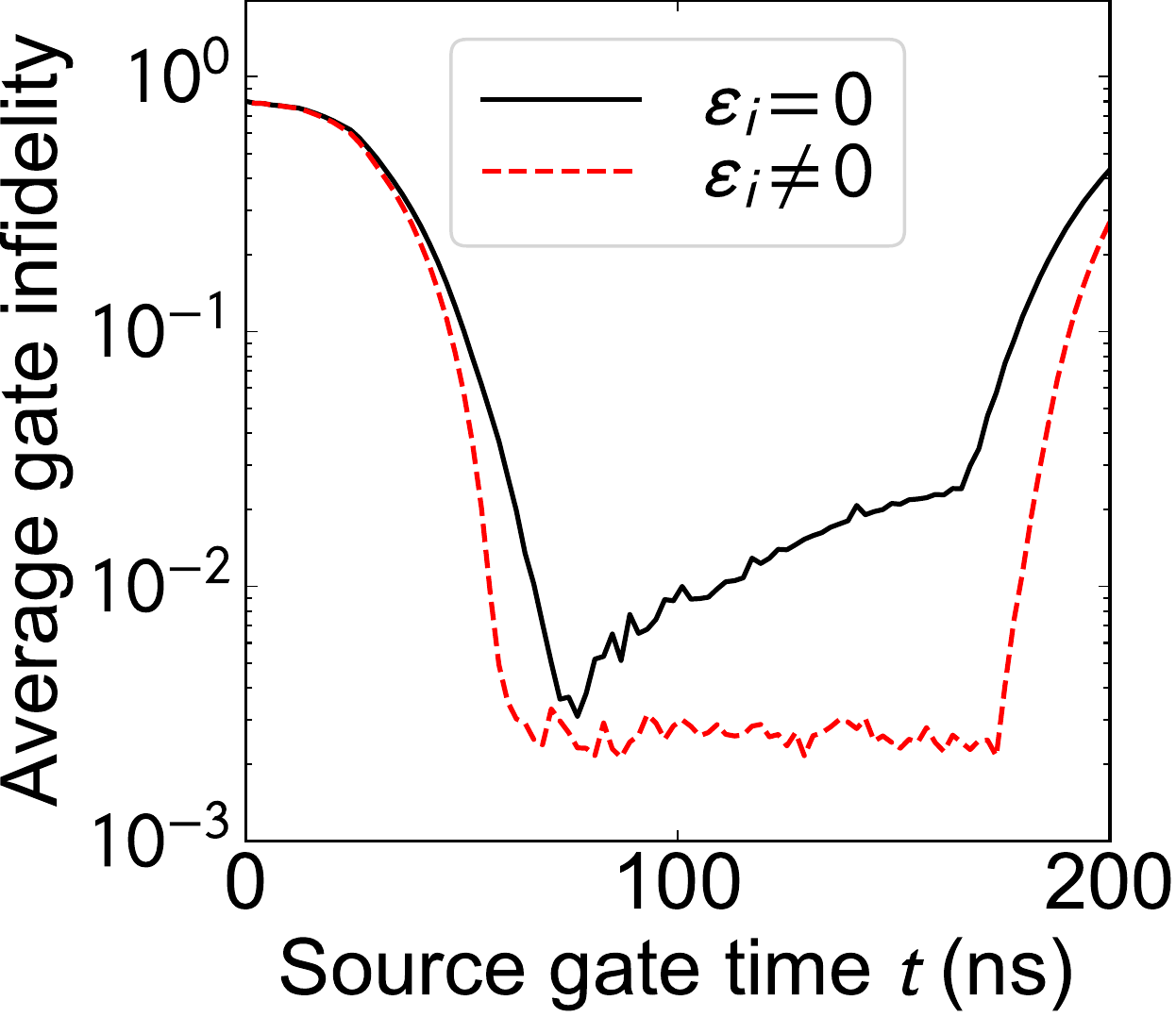}
			\caption{Average gate infidelities of the synthesized four-qubit syndrome extraction as a function of the gate timel $t$ of the source gates. The black solid curve is for $\varepsilon_i=0$ and the red dashed for $\varepsilon_i \ne 0$ (see Table\,\ref{params}). }
			\label{fit_syndrome}
		\end{center}
	\end{figure}

	In the numerical simulations, we choose parameters shown in Table\,\ref{params}, where the detunings $\delta_i$ and the coupling strengths $g_i$ are chosen as similar values to the previous example with additional random fluctuations.
	We simulate the cases with and without crosstalk. In the latter case, the amplitude attenuation $\varepsilon_i$ and the phase delay $\phi_i$ are randomly chosen.
	Drive amplitude $\Omega_i$ is tuned by the above-mentioned concatenated optimization with the gate time of $t=75\ \rm{ns}$. $\Omega_i$ converges to $\Omega_i^{(0)}$ without crosstalk and $\Omega_i^{(\varepsilon_i)}$ with crosstalk. The AGI of the four-qubit syndrome extraction is plotted in Fig.\,\ref{fit_syndrome} as a function of $t$.
	\begin{table}[t]
		\caption{Set of parameters used in the numerical simulation of VQGO for the four-qubit syndrome extraction.}
		\label{params}
		\begin{ruledtabular}
			\begin{tabular}{c|cccccc}
				& $\delta_i\left(\rm{MHz}\right)$ & $g_i\left(\rm{MHz}\right)$ & $\varepsilon_i$ & $\phi_i/\pi$ & $\Omega_i^{(0)}\left(\rm{MHz}\right)$ & $\Omega_i^{(\varepsilon_i)}\left(\rm{MHz}\right)$\\ \hline
				$\rm{Q_1}$ & 211 & 5.0 & 0.1 & 0.2 & 74.5 & 95.9 \\
				$\rm{Q_2}$ & 223 & 5.7 & 0.3 & 1.4 & 79.1 & 85.6 \\
				$\rm{Q_3}$ & 236 & 5.3 & 0.7 & 1.0 & 114  & 106 \\
				$\rm{Q_4}$ & 248 & 5.4 & 0.2 & 0.6 & 115  & 105 \\
			\end{tabular}
		\end{ruledtabular}
	\end{table}
	
	The AGI obtained at $t=75\,\rm{ns}$ is $0.0037$ without crosstalk and $0.0027$ with crosstalk.
	This demonstrates the four-qubit syndrome extraction using the source gates only twice and taking only $150\ \rm{ns}$ excluding the time for single-qubit gates.
	In contrast, a syndrome extraction implemented with sequential CNOT gates would take more than $600\ \rm{ns}$. Moreover, to match the AGI achieved with VQGO, the AGI of the each CNOT gate would have to be less than $0.001$ on average.
	
	
	{\it Conclusion.}
	We proposed VQGO as a HQC algorithm for gate optimization.
	VQGO is easy to implement and suitable for the current experimental situation in which the gate fidelities of single-qubit gates are much better than those of multi-qubit gates.
	This method also has an advantage that we do not need to generate arbitrarily shaped pulses. Therefore, VQGO provides a practical approach for near-term quantum devices. 
	
	In real devices, there exist stochastic gate errors in addition to the coherent errors discussed here. We expect that VQGO also works properly under such errors by finding the decoherence-free subspace automatically.
	On the other hand, if there are state-preparation and measurement (SPAM) errors, the minimization of the cost function does not necessarily converge to the optimal AGI.
	In such a case, we can circumvent the effect of SPAM errors by error mitigations~\cite{temme2017error,endo2018practical,li2017efficient} or by self-consistent quantum tomography~\cite{merkel2013self,blume2013robust,sugiyama2018reliable}.
	%
	
	VQE is the optimization method for quantum states, while VQGO is that for quantum processes.
	The relation between VQE and VQGO are similar to that between quantum state tomography and quantum process tomography. The extension from VQE to VQGO proposed here provides a more general application to realize desired quantum dynamics on a quantum computer.
	
	
	
	\section{Acknowledgements}
	We acknowledge fruitful discussions with T. Sugiyama, S. Kono, S. Tamate, and Y. Tabuchi. This work was supported in part by the Japan Society for the Promotion of Science (JSPS) Grants-in-Aid for Scientific Research (KAKENHI) (No.\,26220601), and the Japan Science and Technology Agency (JST) Exploratory Research for Advanced Technology (ERATO) (No.\,JPMJER1601). The numerical simulations were performed with Quantum Toolbox in Python (QuTiP)~\cite{johansson2012qutip}, Open Source Scientific Tools for Python (SciPy)~\cite{scipy}, and TensorFlow~\cite{tensorflow2015-whitepaper}.
	
	\pagebreak
	\begin{widetext}
		\begin{center}		
			\textbf{\large Supplementary Information for ``Variational Quantum Gate Optimization''}
		\end{center}
	\end{widetext}
	
	\setcounter{equation}{0}
	\setcounter{figure}{0}
	\setcounter{table}{0}
	\setcounter{page}{1}
	\makeatletter
	\renewcommand{\theequation}{S\arabic{equation}}
	\renewcommand{\thefigure}{S\arabic{figure}}
	
	\section{Direct Fidelity Estimation}
	In this section, we denote $\ket{\pm}$, $\ket{i\pm}$, and $\ket{0,1}$ as the eigenvectors of the $\sigma_x$, $\sigma_y$, and $\sigma_z$ with eigenvalues $\pm1$, respectively.
	In quantum process tomography, the initial quantum state and measurement basis are prepared from all the combinations of $\{\ket{0},\ket{1},\ket{+},\ket{i+}\}^{\otimes n} \otimes \{\sigma_x,\sigma_y,\sigma_z\}^{\otimes n}$. Thus, the number of integrated measurements amounts to $12^n$. From measured values, the quantum gate is reconstructed as $4^n\times 4^n$ complex matrix called a process matrix.
	
	On the other hands, in direct fidelity estimation (DFE)~\cite{PhysRevLett.106.230501}, we characterize not the full details of the quantum gate but the average gate fidelity (AGF) between the experimental quantum gate $\mathcal{E}$ and an ideal target gate $\mathcal{U}$, which is defined as follows,
	\begin{align}
	F_{\rm{ave}}\left(\mathcal{U},\mathcal{E}\right)
	&=\int \mathrm{Tr}\left[\mathcal{U}\left(\psi\right)\mathcal{E}\left(\psi\right)\right] d\mu_\psi\\
	&=\frac{1}{D+1}\left\{\frac{1}{D}\mathrm{Tr}\left[\mathcal{U}^\dagger \mathcal{E}\right]+1\right\},
	\end{align}
	where $D$ is the dimention of the gates.
	To characterize the AGF, we can use the following relation,
	\begin{align}
	&\mathrm{Tr}\left[\mathcal{U}^\dagger \mathcal{E}\right]=\sum_{i,j}R\left(\mathcal{U}\right)_{ij}R\left(\mathcal{E}\right)_{ij}\\
	&\left(R\left(\mathcal{U}\right)_{ij}=\frac{1}{D}\mathrm{Tr}\left[\sigma_i\mathcal{U}\left(\sigma_j\right)\right]\right),
	\end{align}
	where $R$ is Pauli transfer matrix (PTM), and $i$ and $j$ are the labels for $n$-qubit Pauli operators from $1$ to $4^n$.
	Then, we can denote the AGF as follows,
	\begin{align}
	F_{\rm{ave}}\left(\mathcal{U},\mathcal{E}\right)
	&=\frac{1}{D+1}\left\{\frac{1}{D}\sum_{i,j}R\left(\mathcal{U}\right)_{ij}R\left(\mathcal{E}\right)_{ij}+1\right\}.
	\end{align}
	Therefore, to estimate the AGF, we need to characterize the components of the PTM.
	The PTM of the ideal target gate can be calculated on a classical computer, and the PTM of the experimental gate can be characterized by measurements as follows,
	\begin{align}
	R\left(\mathcal{U}\right)_{ij}
	&=\frac{1}{D}\, \mathrm{Tr} \!\left[\sigma_i \, \mathcal{U} \!\left(\sum_k \lambda_{jk}\ket{\psi}_{jk}\bra{\psi}_{jk}\right)\right]\\
	&=\frac{1}{D}\sum_{k=1}^{2^n} \lambda_{jk}\, \mathrm{Tr} \!\left[\sigma_i \, \mathcal{U} \!\left(\ket{\psi}_{jk}\bra{\psi}_{jk}\right)\right],
	\end{align}
	where $\ket{\psi}_{jk}$ and $\lambda_{jk}$ are the eigenvector and the eigenvalue of the Pauli operator $\sigma_j$, respectively.
	
	Therefore, in DFE, the set of initial quantum state and measurement basis  for efficient estimation of the AGF depends on the PTM of the target gate. The set is chosen stochastically from all combination of $\{\ket{0},\ket{1},\ket{+},\ket{-},\ket{i+},\ket{i-}\}^{\otimes n} \otimes \{I,\sigma_x,\sigma_y,\sigma_z\}^{\otimes n}$. Probability to choose the set $P_{ij}$ is determined by the value of the PTM of the target gate as follows,
	\begin{align}
	P_{ij} = \frac{1}{D^2}R\left(\mathcal{U}\right)_{ij}^2.
	\end{align}
	The AGF is estimated from the measured values under a set stochastically chosen, i.e.,
	\begin{align}
	F_{\rm{ave}}\left(\mathcal{U},\mathcal{E}\right)
	&=\frac{1}{D+1}\left\{D\sum_{i,j}P_{ij}\left[\frac{R\left(\mathcal{E}\right)_{ij}}{R\left(\mathcal{U}\right)_{ij}}\right]+1\right\}.
	\end{align}
	DFE has a finite failure probability and an estimation error due to the finite number of the integrated measurements. With $\varepsilon^{-2} \delta^{-1}$-time single-shot measurements which are executed on the set of initial quantum states and measurement bases chosen stochastically, DFE estimates AGF with a success probability of $1-\varepsilon$, with accuracy of $\delta$.
	
	The PTM of a Clifford gate only have $-1$ or $1$ in each row or line and all remained components are $0$.
	Therefore, the set is chosen from $8^n$ instances with equal probability.  In this letter, we adopt full measurements on all the sets in which the PTM of the target gate is not $0$.
	
	\section{Representation power of parametrized quantum circuit}
	\subsection{Cartan decomposition}
	Cartan decomposition is a decomposition of a semi-simple Lie group~\cite{helgason2001differential}. The simplest notation of Cartan decomposition of SU(4) is that an arbitrary gate in SU(4) can be decomposed into the degrees of freedom in SU(2) and Cartan subgroups as follows,
	\begin{align}
	U_2 = U_1^{(1)}\otimes U_1^{(2)} \exp \!\left(i\sum_{i=x,y,z}C_{ii}\sigma_i\otimes \sigma_i\right) U_1^{(3)} \otimes U_1^{(4)},
	\end{align}
	where $U_2$ is a gate in SU(4), $U_1^{(j)}$~($1\le j \le 4$) is a gate in SU(2), and we call $C_{ii}\in[0,\pi/4]$ Cartan coefficients.
	In VQGO, all the multi-qubit source gates are multiplied by single-qubit gates from both sides. Therefore, only the Cartan coefficients of the source gates affect the performance of the parametrized quantum circuits for $n = 2$.
	
	\subsection{Entangling power}
	Entangling power is a quantitative indicator on the ability of a two-qubit gate to generate entanglement~\cite{zanardi2001entanglement}. The entangling power is written as follows,
	\begin{align}
	e_p(U)&\equiv \iint E\left(U\ket{\psi_A}\otimes\ket{\psi_B}\right) d\mu_{\psi_A}d\mu_{\psi_B}\\
	E(\ket{\Psi})&=1-\mathrm{Tr}_A\left[\mathrm{Tr}_B\left[\ket{\Psi}\bra{\Psi}\right]^2\right].
	\end{align}
	It is known that the entangling power is given analytically with the following equation without averaging for an arbitrary tensor product states~\cite{zanardi2001entanglement},
	\begin{align}
	e_p(U)=\left(\frac{D}{D+1}\right)^2 \left[E(U)+E(US)-E(S)\right],
	\end{align}
	where $D$ is the dimention of the Hilbert space, $S$ is the SWAP gate, and $E(U)$ is the linearized entanglement entropy of a unitary gate $U$ defined as follows,
	\begin{align}
	U&=\sum_{i=1}^{D^2} \sqrt{\lambda_i}A_i\otimes B_i,\\
	E(U)&=1-\frac{1}{D^4}\sum_{i=1}^{D^2} \lambda_i^2.
	\end{align}
	Here, $A_i$ and $B_i$ are orthonormal operators and the bases for the Schmid decomposition of $U$, and $\lambda_i$ is their coefficient. $E(U)$ is also called operator entanglement~\cite{nielsen2003quantum}. 
	
	\subsection{Representation power of parametrized quantum circuit for SU$(4)$}
\begin{figure*}[tp]
		\begin{center}
			\includegraphics[width=15 cm]{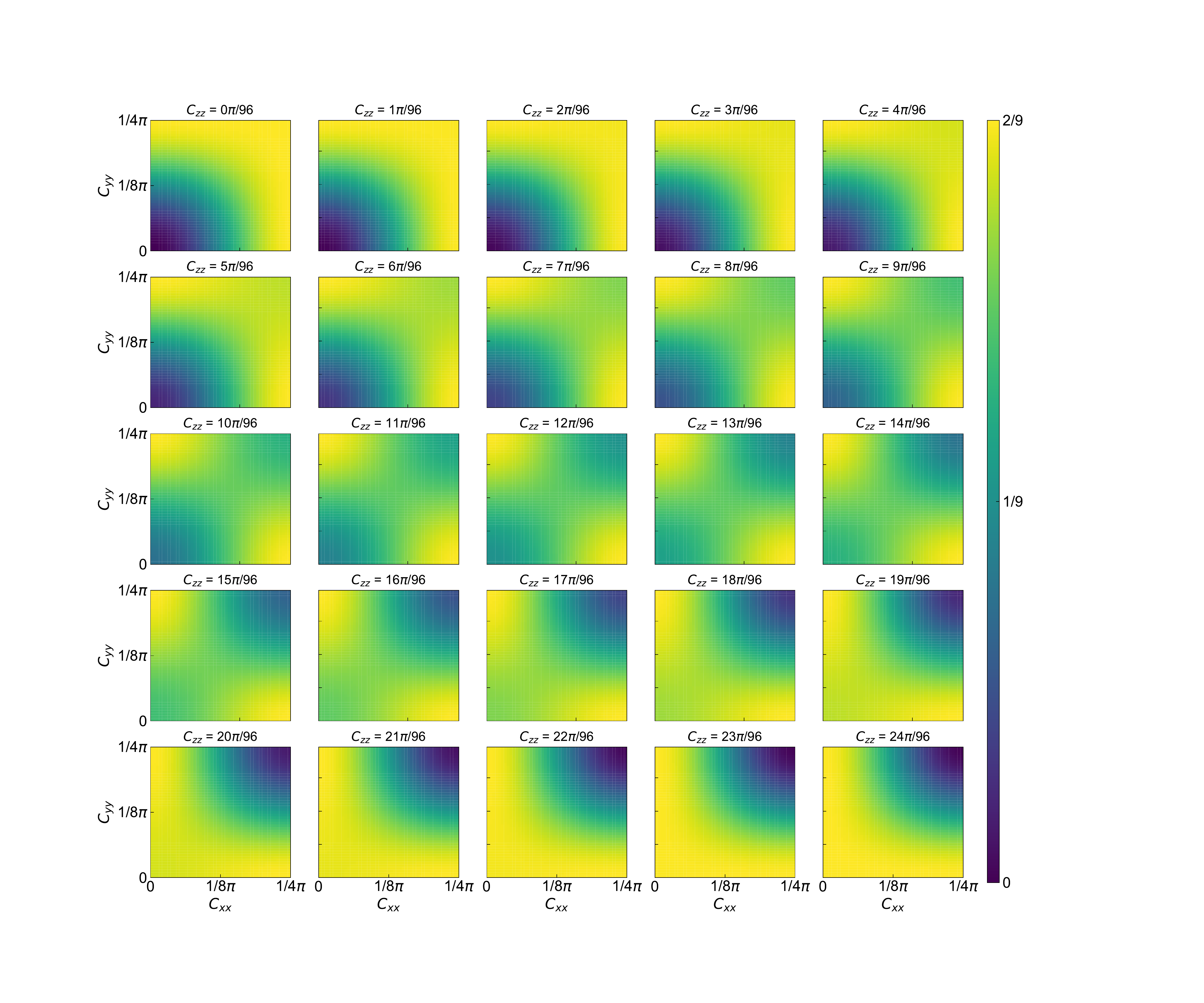}
			\caption{Entangling power of a two-qubit gate in SU(4) plotted in the Cartan space.}
			\label{cartan_ep}
			
			\includegraphics[width=15 cm]{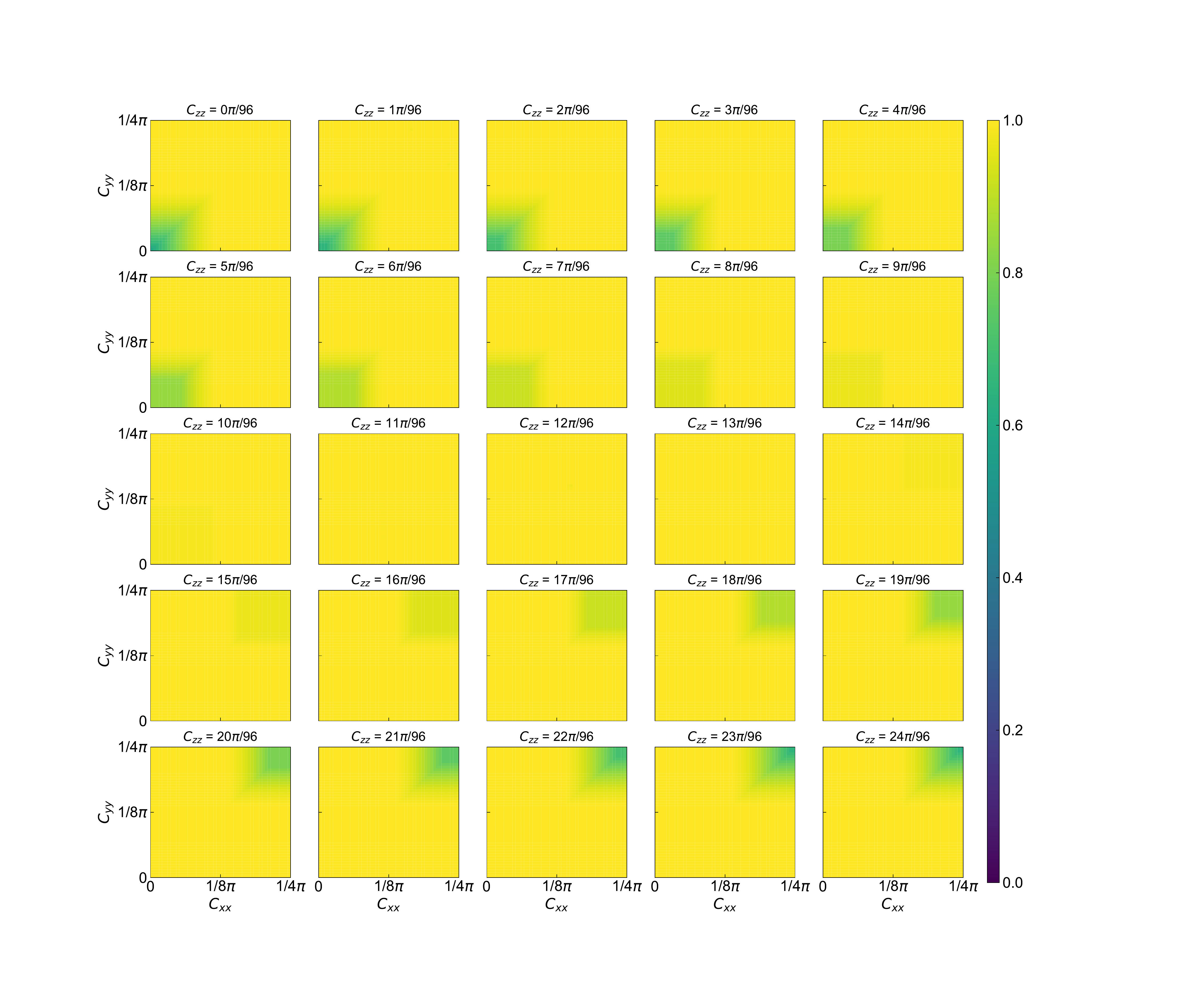}
			\caption{Average gate fidelity between the optimized parametrized quantum circuit with two source gates and a CNOT gate plotted in the Cartan space.}
			\label{cartan_agf}
			
		\end{center}
	\end{figure*}

	The entangling power of two-qubit gates are plotted in Fig.\,\ref{cartan_ep} as a function of the Cartan coefficients.
	The entangling power decreases when the coefficients approach $(0,0,0)$ or $(\pi/4,\pi/4,\pi/4)$, which correspond to an identity gate and a SWAP gate, respectively.
	The average gate fidelities (AGF) between the parametrized quantum circuits with depth $d=2$ and a CNOT gate are plotted in Fig.\,\ref{cartan_agf} as a function of the Cartan coefficients of the source gate.
	The AGF converges to almost unity unless the condition $\left(C_{jj}>\frac{\pi}{8}\ \left(\forall j \in (x,y,z)\right)\right)\vee \left(C_{jj}<\frac{\pi}{8}\ \left(\forall j \in (x,y,z)\right)\right)$ are met.
	The AGF decreases when the coefficients approach $(0,0,0)$ or $(\pi/4,\pi/4,\pi/4)$.
	
	\clearpage
	\bibliographystyle{apsrev4-1}
	\bibliography{cite}

\begin{thebibliography}{79}%
\makeatletter
\providecommand \@ifxundefined [1]{%
 \@ifx{#1\undefined}
}%
\providecommand \@ifnum [1]{%
 \ifnum #1\expandafter \@firstoftwo
 \else \expandafter \@secondoftwo
 \fi
}%
\providecommand \@ifx [1]{%
 \ifx #1\expandafter \@firstoftwo
 \else \expandafter \@secondoftwo
 \fi
}%
\providecommand \natexlab [1]{#1}%
\providecommand \enquote  [1]{``#1''}%
\providecommand \bibnamefont  [1]{#1}%
\providecommand \bibfnamefont [1]{#1}%
\providecommand \citenamefont [1]{#1}%
\providecommand \href@noop [0]{\@secondoftwo}%
\providecommand \href [0]{\begingroup \@sanitize@url \@href}%
\providecommand \@href[1]{\@@startlink{#1}\@@href}%
\providecommand \@@href[1]{\endgroup#1\@@endlink}%
\providecommand \@sanitize@url [0]{\catcode `\\12\catcode `\$12\catcode
  `\&12\catcode `\#12\catcode `\^12\catcode `\_12\catcode `\%12\relax}%
\providecommand \@@startlink[1]{}%
\providecommand \@@endlink[0]{}%
\providecommand \url  [0]{\begingroup\@sanitize@url \@url }%
\providecommand \@url [1]{\endgroup\@href {#1}{\urlprefix }}%
\providecommand \urlprefix  [0]{URL }%
\providecommand \Eprint [0]{\href }%
\providecommand \doibase [0]{http://dx.doi.org/}%
\providecommand \selectlanguage [0]{\@gobble}%
\providecommand \bibinfo  [0]{\@secondoftwo}%
\providecommand \bibfield  [0]{\@secondoftwo}%
\providecommand \translation [1]{[#1]}%
\providecommand \BibitemOpen [0]{}%
\providecommand \bibitemStop [0]{}%
\providecommand \bibitemNoStop [0]{.\EOS\space}%
\providecommand \EOS [0]{\spacefactor3000\relax}%
\providecommand \BibitemShut  [1]{\csname bibitem#1\endcsname}%
\let\auto@bib@innerbib\@empty
\bibitem [{\citenamefont {Preskill}(2018)}]{preskill2018quantum}%
  \BibitemOpen
  \bibfield  {author} {\bibinfo {author} {\bibfnamefont {J.}~\bibnamefont
  {Preskill}},\ }\href@noop {} {\bibfield  {journal} {\bibinfo  {journal}
  {arXiv preprint arXiv:1801.00862}\ } (\bibinfo {year} {2018})}\BibitemShut
  {NoStop}%
\bibitem [{\citenamefont {Peruzzo}\ \emph {et~al.}(2014)\citenamefont
  {Peruzzo}, \citenamefont {McClean}, \citenamefont {Shadbolt}, \citenamefont
  {Yung}, \citenamefont {Zhou}, \citenamefont {Love}, \citenamefont
  {Aspuru-Guzik},\ and\ \citenamefont {O'brien}}]{peruzzo2014variational}%
  \BibitemOpen
  \bibfield  {author} {\bibinfo {author} {\bibfnamefont {A.}~\bibnamefont
  {Peruzzo}}, \bibinfo {author} {\bibfnamefont {J.}~\bibnamefont {McClean}},
  \bibinfo {author} {\bibfnamefont {P.}~\bibnamefont {Shadbolt}}, \bibinfo
  {author} {\bibfnamefont {M.-H.}\ \bibnamefont {Yung}}, \bibinfo {author}
  {\bibfnamefont {X.-Q.}\ \bibnamefont {Zhou}}, \bibinfo {author}
  {\bibfnamefont {P.~J.}\ \bibnamefont {Love}}, \bibinfo {author}
  {\bibfnamefont {A.}~\bibnamefont {Aspuru-Guzik}}, \ and\ \bibinfo {author}
  {\bibfnamefont {J.~L.}\ \bibnamefont {O'brien}},\ }\href@noop {} {\bibfield
  {journal} {\bibinfo  {journal} {Nature Communications}\ }\textbf {\bibinfo
  {volume} {5}},\ \bibinfo {pages} {4213} (\bibinfo {year} {2014})}\BibitemShut
  {NoStop}%
\bibitem [{\citenamefont {Kandala}\ \emph {et~al.}(2017)\citenamefont
  {Kandala}, \citenamefont {Mezzacapo}, \citenamefont {Temme}, \citenamefont
  {Takita}, \citenamefont {Brink}, \citenamefont {Chow},\ and\ \citenamefont
  {Gambetta}}]{kandala2017hardware}%
  \BibitemOpen
  \bibfield  {author} {\bibinfo {author} {\bibfnamefont {A.}~\bibnamefont
  {Kandala}}, \bibinfo {author} {\bibfnamefont {A.}~\bibnamefont {Mezzacapo}},
  \bibinfo {author} {\bibfnamefont {K.}~\bibnamefont {Temme}}, \bibinfo
  {author} {\bibfnamefont {M.}~\bibnamefont {Takita}}, \bibinfo {author}
  {\bibfnamefont {M.}~\bibnamefont {Brink}}, \bibinfo {author} {\bibfnamefont
  {J.~M.}\ \bibnamefont {Chow}}, \ and\ \bibinfo {author} {\bibfnamefont
  {J.~M.}\ \bibnamefont {Gambetta}},\ }\href@noop {} {\bibfield  {journal}
  {\bibinfo  {journal} {Nature}\ }\textbf {\bibinfo {volume} {549}},\ \bibinfo
  {pages} {242} (\bibinfo {year} {2017})}\BibitemShut {NoStop}%
\bibitem [{\citenamefont {Colless}\ \emph {et~al.}(2018)\citenamefont
  {Colless}, \citenamefont {Ramasesh}, \citenamefont {Dahlen}, \citenamefont
  {Blok}, \citenamefont {Kimchi-Schwartz}, \citenamefont {McClean},
  \citenamefont {Carter}, \citenamefont {De~Jong},\ and\ \citenamefont
  {Siddiqi}}]{colless2018computation}%
  \BibitemOpen
  \bibfield  {author} {\bibinfo {author} {\bibfnamefont {J.}~\bibnamefont
  {Colless}}, \bibinfo {author} {\bibfnamefont {V.}~\bibnamefont {Ramasesh}},
  \bibinfo {author} {\bibfnamefont {D.}~\bibnamefont {Dahlen}}, \bibinfo
  {author} {\bibfnamefont {M.}~\bibnamefont {Blok}}, \bibinfo {author}
  {\bibfnamefont {M.}~\bibnamefont {Kimchi-Schwartz}}, \bibinfo {author}
  {\bibfnamefont {J.}~\bibnamefont {McClean}}, \bibinfo {author} {\bibfnamefont
  {J.}~\bibnamefont {Carter}}, \bibinfo {author} {\bibfnamefont
  {W.}~\bibnamefont {De~Jong}}, \ and\ \bibinfo {author} {\bibfnamefont
  {I.}~\bibnamefont {Siddiqi}},\ }\href@noop {} {\bibfield  {journal} {\bibinfo
   {journal} {Physical Review X}\ }\textbf {\bibinfo {volume} {8}},\ \bibinfo
  {pages} {011021} (\bibinfo {year} {2018})}\BibitemShut {NoStop}%
\bibitem [{\citenamefont {Farhi}\ \emph {et~al.}(2014)\citenamefont {Farhi},
  \citenamefont {Goldstone},\ and\ \citenamefont {Gutmann}}]{farhi2014quantum}%
  \BibitemOpen
  \bibfield  {author} {\bibinfo {author} {\bibfnamefont {E.}~\bibnamefont
  {Farhi}}, \bibinfo {author} {\bibfnamefont {J.}~\bibnamefont {Goldstone}}, \
  and\ \bibinfo {author} {\bibfnamefont {S.}~\bibnamefont {Gutmann}},\
  }\href@noop {} {\bibfield  {journal} {\bibinfo  {journal} {arXiv preprint
  arXiv:1411.4028}\ } (\bibinfo {year} {2014})}\BibitemShut {NoStop}%
\bibitem [{\citenamefont {Otterbach}\ \emph {et~al.}(2017)\citenamefont
  {Otterbach}, \citenamefont {Manenti}, \citenamefont {Alidoust}, \citenamefont
  {Bestwick}, \citenamefont {Block}, \citenamefont {Bloom}, \citenamefont
  {Caldwell}, \citenamefont {Didier}, \citenamefont {Fried}, \citenamefont
  {Hong}, \citenamefont {Karalekas}, \citenamefont {Osborn}, \citenamefont
  {Papageorge}, \citenamefont {Peterson}, \citenamefont {Prawiroatmodjo},
  \citenamefont {Rubin}, \citenamefont {Ryan}, \citenamefont {Scarabelli},
  \citenamefont {Scheer}, \citenamefont {Sete}, \citenamefont {Sivarajah},
  \citenamefont {Smith}, \citenamefont {Staley}, \citenamefont {Tezak},
  \citenamefont {Zeng}, \citenamefont {Hudson}, \citenamefont {Johnson},
  \citenamefont {Reagor}, \citenamefont {da~Silva},\ and\ \citenamefont
  {Rigetti}}]{otterbach2017unsupervised}%
  \BibitemOpen
  \bibfield  {author} {\bibinfo {author} {\bibfnamefont {J.~S.}\ \bibnamefont
  {Otterbach}}, \bibinfo {author} {\bibfnamefont {R.}~\bibnamefont {Manenti}},
  \bibinfo {author} {\bibfnamefont {N.}~\bibnamefont {Alidoust}}, \bibinfo
  {author} {\bibfnamefont {A.}~\bibnamefont {Bestwick}}, \bibinfo {author}
  {\bibfnamefont {M.}~\bibnamefont {Block}}, \bibinfo {author} {\bibfnamefont
  {B.}~\bibnamefont {Bloom}}, \bibinfo {author} {\bibfnamefont
  {S.}~\bibnamefont {Caldwell}}, \bibinfo {author} {\bibfnamefont
  {N.}~\bibnamefont {Didier}}, \bibinfo {author} {\bibfnamefont {E.~S.}\
  \bibnamefont {Fried}}, \bibinfo {author} {\bibfnamefont {S.}~\bibnamefont
  {Hong}}, \bibinfo {author} {\bibfnamefont {P.}~\bibnamefont {Karalekas}},
  \bibinfo {author} {\bibfnamefont {C.~B.}\ \bibnamefont {Osborn}}, \bibinfo
  {author} {\bibfnamefont {A.}~\bibnamefont {Papageorge}}, \bibinfo {author}
  {\bibfnamefont {E.~C.}\ \bibnamefont {Peterson}}, \bibinfo {author}
  {\bibfnamefont {G.}~\bibnamefont {Prawiroatmodjo}}, \bibinfo {author}
  {\bibfnamefont {N.}~\bibnamefont {Rubin}}, \bibinfo {author} {\bibfnamefont
  {C.~A.}\ \bibnamefont {Ryan}}, \bibinfo {author} {\bibfnamefont
  {D.}~\bibnamefont {Scarabelli}}, \bibinfo {author} {\bibfnamefont
  {M.}~\bibnamefont {Scheer}}, \bibinfo {author} {\bibfnamefont {E.~A.}\
  \bibnamefont {Sete}}, \bibinfo {author} {\bibfnamefont {P.}~\bibnamefont
  {Sivarajah}}, \bibinfo {author} {\bibfnamefont {R.~S.}\ \bibnamefont
  {Smith}}, \bibinfo {author} {\bibfnamefont {A.}~\bibnamefont {Staley}},
  \bibinfo {author} {\bibfnamefont {N.}~\bibnamefont {Tezak}}, \bibinfo
  {author} {\bibfnamefont {W.~J.}\ \bibnamefont {Zeng}}, \bibinfo {author}
  {\bibfnamefont {A.}~\bibnamefont {Hudson}}, \bibinfo {author} {\bibfnamefont
  {B.~R.}\ \bibnamefont {Johnson}}, \bibinfo {author} {\bibfnamefont
  {M.}~\bibnamefont {Reagor}}, \bibinfo {author} {\bibfnamefont {M.~P.}\
  \bibnamefont {da~Silva}}, \ and\ \bibinfo {author} {\bibfnamefont
  {C.}~\bibnamefont {Rigetti}},\ }\href@noop {} {\bibfield  {journal} {\bibinfo
   {journal} {arXiv preprint arXiv:1712.05771}\ } (\bibinfo {year}
  {2017})}\BibitemShut {NoStop}%
\bibitem [{\citenamefont {Mitarai}\ \emph {et~al.}(2018)\citenamefont
  {Mitarai}, \citenamefont {Negoro}, \citenamefont {Kitagawa},\ and\
  \citenamefont {Fujii}}]{mitarai2018quantum}%
  \BibitemOpen
  \bibfield  {author} {\bibinfo {author} {\bibfnamefont {K.}~\bibnamefont
  {Mitarai}}, \bibinfo {author} {\bibfnamefont {M.}~\bibnamefont {Negoro}},
  \bibinfo {author} {\bibfnamefont {M.}~\bibnamefont {Kitagawa}}, \ and\
  \bibinfo {author} {\bibfnamefont {K.}~\bibnamefont {Fujii}},\ }\href
  {\doibase 10.1103/PhysRevA.98.032309} {\bibfield  {journal} {\bibinfo
  {journal} {Physical Review A}\ }\textbf {\bibinfo {volume} {98}},\ \bibinfo
  {pages} {032309} (\bibinfo {year} {2018})}\BibitemShut {NoStop}%
\bibitem [{\citenamefont {Havlicek}\ \emph {et~al.}(2018)\citenamefont
  {Havlicek}, \citenamefont {D.~C~^^cc^^81orcoles}, \citenamefont {Temme},
  \citenamefont {W.~Harrow}, \citenamefont {Kandala}, \citenamefont {M.~Chow},\
  and\ \citenamefont {Gambetta}}]{vojtech2018quantum}%
  \BibitemOpen
  \bibfield  {author} {\bibinfo {author} {\bibfnamefont {V.}~\bibnamefont
  {Havlicek}}, \bibinfo {author} {\bibfnamefont {A.}~\bibnamefont
  {D.~C~^^cc^^81orcoles}}, \bibinfo {author} {\bibfnamefont {K.}~\bibnamefont
  {Temme}}, \bibinfo {author} {\bibfnamefont {A.}~\bibnamefont {W.~Harrow}},
  \bibinfo {author} {\bibfnamefont {A.}~\bibnamefont {Kandala}}, \bibinfo
  {author} {\bibfnamefont {J.}~\bibnamefont {M.~Chow}}, \ and\ \bibinfo
  {author} {\bibfnamefont {J.~M.}\ \bibnamefont {Gambetta}},\ }\href@noop {}
  {\bibfield  {journal} {\bibinfo  {journal} {arXiv preprint arXiv:1804.11326}\
  } (\bibinfo {year} {2018})}\BibitemShut {NoStop}%
\bibitem [{\citenamefont {Wecker}\ \emph {et~al.}(2015)\citenamefont {Wecker},
  \citenamefont {Hastings}, \citenamefont {Wiebe}, \citenamefont {Clark},
  \citenamefont {Nayak},\ and\ \citenamefont {Troyer}}]{wecker2015solving}%
  \BibitemOpen
  \bibfield  {author} {\bibinfo {author} {\bibfnamefont {D.}~\bibnamefont
  {Wecker}}, \bibinfo {author} {\bibfnamefont {M.~B.}\ \bibnamefont
  {Hastings}}, \bibinfo {author} {\bibfnamefont {N.}~\bibnamefont {Wiebe}},
  \bibinfo {author} {\bibfnamefont {B.~K.}\ \bibnamefont {Clark}}, \bibinfo
  {author} {\bibfnamefont {C.}~\bibnamefont {Nayak}}, \ and\ \bibinfo {author}
  {\bibfnamefont {M.}~\bibnamefont {Troyer}},\ }\href@noop {} {\bibfield
  {journal} {\bibinfo  {journal} {Physical Review A}\ }\textbf {\bibinfo
  {volume} {92}},\ \bibinfo {pages} {062318} (\bibinfo {year}
  {2015})}\BibitemShut {NoStop}%
\bibitem [{\citenamefont {McClean}\ \emph {et~al.}(2016)\citenamefont
  {McClean}, \citenamefont {Romero}, \citenamefont {Babbush},\ and\
  \citenamefont {Aspuru-Guzik}}]{mcclean2016theory}%
  \BibitemOpen
  \bibfield  {author} {\bibinfo {author} {\bibfnamefont {J.~R.}\ \bibnamefont
  {McClean}}, \bibinfo {author} {\bibfnamefont {J.}~\bibnamefont {Romero}},
  \bibinfo {author} {\bibfnamefont {R.}~\bibnamefont {Babbush}}, \ and\
  \bibinfo {author} {\bibfnamefont {A.}~\bibnamefont {Aspuru-Guzik}},\
  }\href@noop {} {\bibfield  {journal} {\bibinfo  {journal} {New Journal of
  Physics}\ }\textbf {\bibinfo {volume} {18}},\ \bibinfo {pages} {023023}
  (\bibinfo {year} {2016})}\BibitemShut {NoStop}%
\bibitem [{\citenamefont {Wang}\ \emph {et~al.}(2017)\citenamefont {Wang},
  \citenamefont {Paesani}, \citenamefont {Santagati}, \citenamefont {Knauer},
  \citenamefont {Gentile}, \citenamefont {Wiebe}, \citenamefont {Petruzzella},
  \citenamefont {O'Brien}, \citenamefont {Rarity}, \citenamefont {Laing},\ and\
  \citenamefont {Thompson}}]{wang2017experimental}%
  \BibitemOpen
  \bibfield  {author} {\bibinfo {author} {\bibfnamefont {J.}~\bibnamefont
  {Wang}}, \bibinfo {author} {\bibfnamefont {S.}~\bibnamefont {Paesani}},
  \bibinfo {author} {\bibfnamefont {R.}~\bibnamefont {Santagati}}, \bibinfo
  {author} {\bibfnamefont {S.}~\bibnamefont {Knauer}}, \bibinfo {author}
  {\bibfnamefont {A.~A.}\ \bibnamefont {Gentile}}, \bibinfo {author}
  {\bibfnamefont {N.}~\bibnamefont {Wiebe}}, \bibinfo {author} {\bibfnamefont
  {M.}~\bibnamefont {Petruzzella}}, \bibinfo {author} {\bibfnamefont {J.~L.}\
  \bibnamefont {O'Brien}}, \bibinfo {author} {\bibfnamefont {J.~G.}\
  \bibnamefont {Rarity}}, \bibinfo {author} {\bibfnamefont {A.}~\bibnamefont
  {Laing}}, \ and\ \bibinfo {author} {\bibfnamefont {M.~G.}\ \bibnamefont
  {Thompson}},\ }\href@noop {} {\bibfield  {journal} {\bibinfo  {journal}
  {Nature Physics}\ }\textbf {\bibinfo {volume} {13}},\ \bibinfo {pages} {551}
  (\bibinfo {year} {2017})}\BibitemShut {NoStop}%
\bibitem [{\citenamefont {Guerreschi}\ and\ \citenamefont
  {Smelyanskiy}(2017)}]{guerreschi2017practical}%
  \BibitemOpen
  \bibfield  {author} {\bibinfo {author} {\bibfnamefont {G.~G.}\ \bibnamefont
  {Guerreschi}}\ and\ \bibinfo {author} {\bibfnamefont {M.}~\bibnamefont
  {Smelyanskiy}},\ }\href@noop {} {\bibfield  {journal} {\bibinfo  {journal}
  {arXiv preprint arXiv:1701.01450}\ } (\bibinfo {year} {2017})}\BibitemShut
  {NoStop}%
\bibitem [{\citenamefont {McClean}\ \emph {et~al.}(2017)\citenamefont
  {McClean}, \citenamefont {Kimchi-Schwartz}, \citenamefont {Carter},\ and\
  \citenamefont {de~Jong}}]{mcclean2017hybrid}%
  \BibitemOpen
  \bibfield  {author} {\bibinfo {author} {\bibfnamefont {J.~R.}\ \bibnamefont
  {McClean}}, \bibinfo {author} {\bibfnamefont {M.~E.}\ \bibnamefont
  {Kimchi-Schwartz}}, \bibinfo {author} {\bibfnamefont {J.}~\bibnamefont
  {Carter}}, \ and\ \bibinfo {author} {\bibfnamefont {W.~A.}\ \bibnamefont
  {de~Jong}},\ }\href@noop {} {\bibfield  {journal} {\bibinfo  {journal}
  {Physical Review A}\ }\textbf {\bibinfo {volume} {95}},\ \bibinfo {pages}
  {042308} (\bibinfo {year} {2017})}\BibitemShut {NoStop}%
\bibitem [{\citenamefont {Romero}\ \emph {et~al.}(2018)\citenamefont {Romero},
  \citenamefont {Babbush}, \citenamefont {McClean}, \citenamefont {Hempel},
  \citenamefont {Love},\ and\ \citenamefont
  {Aspuru-Guzik}}]{romero2018strategies}%
  \BibitemOpen
  \bibfield  {author} {\bibinfo {author} {\bibfnamefont {J.}~\bibnamefont
  {Romero}}, \bibinfo {author} {\bibfnamefont {R.}~\bibnamefont {Babbush}},
  \bibinfo {author} {\bibfnamefont {J.}~\bibnamefont {McClean}}, \bibinfo
  {author} {\bibfnamefont {C.}~\bibnamefont {Hempel}}, \bibinfo {author}
  {\bibfnamefont {P.}~\bibnamefont {Love}}, \ and\ \bibinfo {author}
  {\bibfnamefont {A.}~\bibnamefont {Aspuru-Guzik}},\ }\href@noop {} {\bibfield
  {journal} {\bibinfo  {journal} {Quantum Science and Technology}\ } (\bibinfo
  {year} {2018})}\BibitemShut {NoStop}%
\bibitem [{\citenamefont {Johnson}\ \emph {et~al.}(2017)\citenamefont
  {Johnson}, \citenamefont {Romero}, \citenamefont {Olson}, \citenamefont
  {Cao},\ and\ \citenamefont {Aspuru-Guzik}}]{johnson2017qvector}%
  \BibitemOpen
  \bibfield  {author} {\bibinfo {author} {\bibfnamefont {P.~D.}\ \bibnamefont
  {Johnson}}, \bibinfo {author} {\bibfnamefont {J.}~\bibnamefont {Romero}},
  \bibinfo {author} {\bibfnamefont {J.}~\bibnamefont {Olson}}, \bibinfo
  {author} {\bibfnamefont {Y.}~\bibnamefont {Cao}}, \ and\ \bibinfo {author}
  {\bibfnamefont {A.}~\bibnamefont {Aspuru-Guzik}},\ }\href@noop {} {\bibfield
  {journal} {\bibinfo  {journal} {arXiv preprint arXiv:1711.02249}\ } (\bibinfo
  {year} {2017})}\BibitemShut {NoStop}%
\bibitem [{\citenamefont {Khatri}\ \emph {et~al.}(2018)\citenamefont {Khatri},
  \citenamefont {LaRose}, \citenamefont {Poremba}, \citenamefont {Cincio},
  \citenamefont {Sornborger},\ and\ \citenamefont {Coles}}]{khatri2018quantum}%
  \BibitemOpen
  \bibfield  {author} {\bibinfo {author} {\bibfnamefont {S.}~\bibnamefont
  {Khatri}}, \bibinfo {author} {\bibfnamefont {R.}~\bibnamefont {LaRose}},
  \bibinfo {author} {\bibfnamefont {A.}~\bibnamefont {Poremba}}, \bibinfo
  {author} {\bibfnamefont {L.}~\bibnamefont {Cincio}}, \bibinfo {author}
  {\bibfnamefont {A.~T.}\ \bibnamefont {Sornborger}}, \ and\ \bibinfo {author}
  {\bibfnamefont {P.~J.}\ \bibnamefont {Coles}},\ }\href@noop {} {\bibfield
  {journal} {\bibinfo  {journal} {arXiv preprint arXiv:1807.00800}\ } (\bibinfo
  {year} {2018})}\BibitemShut {NoStop}%
\bibitem [{\citenamefont {Takita}\ \emph {et~al.}(2017)\citenamefont {Takita},
  \citenamefont {Cross}, \citenamefont {C{\'o}rcoles}, \citenamefont {Chow},\
  and\ \citenamefont {Gambetta}}]{takita2017experimental}%
  \BibitemOpen
  \bibfield  {author} {\bibinfo {author} {\bibfnamefont {M.}~\bibnamefont
  {Takita}}, \bibinfo {author} {\bibfnamefont {A.~W.}\ \bibnamefont {Cross}},
  \bibinfo {author} {\bibfnamefont {A.}~\bibnamefont {C{\'o}rcoles}}, \bibinfo
  {author} {\bibfnamefont {J.~M.}\ \bibnamefont {Chow}}, \ and\ \bibinfo
  {author} {\bibfnamefont {J.~M.}\ \bibnamefont {Gambetta}},\ }\href@noop {}
  {\bibfield  {journal} {\bibinfo  {journal} {Physical Review Letters}\
  }\textbf {\bibinfo {volume} {119}},\ \bibinfo {pages} {180501} (\bibinfo
  {year} {2017})}\BibitemShut {NoStop}%
\bibitem [{\citenamefont {Kelly}\ \emph {et~al.}(2015)\citenamefont {Kelly},
  \citenamefont {Barends}, \citenamefont {Fowler}, \citenamefont {Megrant},
  \citenamefont {Jeffrey}, \citenamefont {White}, \citenamefont {Sank},
  \citenamefont {Mutus}, \citenamefont {Campbell}, \citenamefont {Chen},
  \citenamefont {Chen}, \citenamefont {Chiaro}, \citenamefont {Dunsworth},
  \citenamefont {Hoi}, \citenamefont {Neill}, \citenamefont {O'Malley},
  \citenamefont {Quintana}, \citenamefont {Roushan}, \citenamefont
  {Vainsencher}, \citenamefont {Wenner}, \citenamefont {Cleland},\ and\
  \citenamefont {Martinis}}]{kelly2015state}%
  \BibitemOpen
  \bibfield  {author} {\bibinfo {author} {\bibfnamefont {J.}~\bibnamefont
  {Kelly}}, \bibinfo {author} {\bibfnamefont {R.}~\bibnamefont {Barends}},
  \bibinfo {author} {\bibfnamefont {A.}~\bibnamefont {Fowler}}, \bibinfo
  {author} {\bibfnamefont {A.}~\bibnamefont {Megrant}}, \bibinfo {author}
  {\bibfnamefont {E.}~\bibnamefont {Jeffrey}}, \bibinfo {author} {\bibfnamefont
  {T.}~\bibnamefont {White}}, \bibinfo {author} {\bibfnamefont
  {D.}~\bibnamefont {Sank}}, \bibinfo {author} {\bibfnamefont {J.}~\bibnamefont
  {Mutus}}, \bibinfo {author} {\bibfnamefont {B.}~\bibnamefont {Campbell}},
  \bibinfo {author} {\bibfnamefont {Y.}~\bibnamefont {Chen}}, \bibinfo {author}
  {\bibfnamefont {Z.}~\bibnamefont {Chen}}, \bibinfo {author} {\bibfnamefont
  {B.}~\bibnamefont {Chiaro}}, \bibinfo {author} {\bibfnamefont
  {A.}~\bibnamefont {Dunsworth}}, \bibinfo {author} {\bibfnamefont {I.~C.}\
  \bibnamefont {Hoi}}, \bibinfo {author} {\bibfnamefont {C.}~\bibnamefont
  {Neill}}, \bibinfo {author} {\bibfnamefont {P.~J.~J.}\ \bibnamefont
  {O'Malley}}, \bibinfo {author} {\bibfnamefont {C.}~\bibnamefont {Quintana}},
  \bibinfo {author} {\bibfnamefont {P.}~\bibnamefont {Roushan}}, \bibinfo
  {author} {\bibfnamefont {A.}~\bibnamefont {Vainsencher}}, \bibinfo {author}
  {\bibfnamefont {J.}~\bibnamefont {Wenner}}, \bibinfo {author} {\bibfnamefont
  {A.~N.}\ \bibnamefont {Cleland}}, \ and\ \bibinfo {author} {\bibfnamefont
  {J.~M.}\ \bibnamefont {Martinis}},\ }\href@noop {} {\bibfield  {journal}
  {\bibinfo  {journal} {Nature}\ }\textbf {\bibinfo {volume} {519}},\ \bibinfo
  {pages} {66} (\bibinfo {year} {2015})}\BibitemShut {NoStop}%
\bibitem [{\citenamefont {C{\'o}rcoles}\ \emph {et~al.}(2015)\citenamefont
  {C{\'o}rcoles}, \citenamefont {Magesan}, \citenamefont {Srinivasan},
  \citenamefont {Cross}, \citenamefont {Steffen}, \citenamefont {Gambetta},\
  and\ \citenamefont {Chow}}]{corcoles2015demonstration}%
  \BibitemOpen
  \bibfield  {author} {\bibinfo {author} {\bibfnamefont {A.~D.}\ \bibnamefont
  {C{\'o}rcoles}}, \bibinfo {author} {\bibfnamefont {E.}~\bibnamefont
  {Magesan}}, \bibinfo {author} {\bibfnamefont {S.~J.}\ \bibnamefont
  {Srinivasan}}, \bibinfo {author} {\bibfnamefont {A.~W.}\ \bibnamefont
  {Cross}}, \bibinfo {author} {\bibfnamefont {M.}~\bibnamefont {Steffen}},
  \bibinfo {author} {\bibfnamefont {J.~M.}\ \bibnamefont {Gambetta}}, \ and\
  \bibinfo {author} {\bibfnamefont {J.~M.}\ \bibnamefont {Chow}},\ }\href@noop
  {} {\bibfield  {journal} {\bibinfo  {journal} {Nature Communications}\
  }\textbf {\bibinfo {volume} {6}},\ \bibinfo {pages} {6979} (\bibinfo {year}
  {2015})}\BibitemShut {NoStop}%
\bibitem [{\citenamefont {Barends}\ \emph {et~al.}(2014)\citenamefont
  {Barends}, \citenamefont {Kelly}, \citenamefont {Megrant}, \citenamefont
  {Veitia}, \citenamefont {Sank}, \citenamefont {Jeffrey}, \citenamefont
  {White}, \citenamefont {Mutus}, \citenamefont {Fowler}, \citenamefont
  {Campbell}, \citenamefont {Chen}, \citenamefont {Chen}, \citenamefont
  {Chiaro}, \citenamefont {Dunsworth}, \citenamefont {Neill}, \citenamefont
  {O'Malley}, \citenamefont {Roushan}, \citenamefont {Vainsencher},
  \citenamefont {Wenner}, \citenamefont {Korotkov}, \citenamefont {Cleland},\
  and\ \citenamefont {Martinis}}]{barends2014logic}%
  \BibitemOpen
  \bibfield  {author} {\bibinfo {author} {\bibfnamefont {R.}~\bibnamefont
  {Barends}}, \bibinfo {author} {\bibfnamefont {J.}~\bibnamefont {Kelly}},
  \bibinfo {author} {\bibfnamefont {A.}~\bibnamefont {Megrant}}, \bibinfo
  {author} {\bibfnamefont {A.}~\bibnamefont {Veitia}}, \bibinfo {author}
  {\bibfnamefont {D.}~\bibnamefont {Sank}}, \bibinfo {author} {\bibfnamefont
  {E.}~\bibnamefont {Jeffrey}}, \bibinfo {author} {\bibfnamefont
  {T.}~\bibnamefont {White}}, \bibinfo {author} {\bibfnamefont
  {J.}~\bibnamefont {Mutus}}, \bibinfo {author} {\bibfnamefont
  {A.}~\bibnamefont {Fowler}}, \bibinfo {author} {\bibfnamefont
  {B.}~\bibnamefont {Campbell}}, \bibinfo {author} {\bibfnamefont
  {Y.}~\bibnamefont {Chen}}, \bibinfo {author} {\bibfnamefont {Z.}~\bibnamefont
  {Chen}}, \bibinfo {author} {\bibfnamefont {B.}~\bibnamefont {Chiaro}},
  \bibinfo {author} {\bibfnamefont {A.}~\bibnamefont {Dunsworth}}, \bibinfo
  {author} {\bibfnamefont {C.}~\bibnamefont {Neill}}, \bibinfo {author}
  {\bibfnamefont {P.}~\bibnamefont {O'Malley}}, \bibinfo {author}
  {\bibfnamefont {P.}~\bibnamefont {Roushan}}, \bibinfo {author} {\bibfnamefont
  {A.}~\bibnamefont {Vainsencher}}, \bibinfo {author} {\bibfnamefont
  {J.}~\bibnamefont {Wenner}}, \bibinfo {author} {\bibfnamefont {A.~N.}\
  \bibnamefont {Korotkov}}, \bibinfo {author} {\bibfnamefont {A.~N.}\
  \bibnamefont {Cleland}}, \ and\ \bibinfo {author} {\bibfnamefont {J.~M.}\
  \bibnamefont {Martinis}},\ }\href@noop {} {\bibfield  {journal} {\bibinfo
  {journal} {arXiv preprint arXiv:1402.4848}\ } (\bibinfo {year}
  {2014})}\BibitemShut {NoStop}%
\bibitem [{\citenamefont {Takita}\ \emph {et~al.}(2016)\citenamefont {Takita},
  \citenamefont {C{\'o}rcoles}, \citenamefont {Magesan}, \citenamefont {Abdo},
  \citenamefont {Brink}, \citenamefont {Cross}, \citenamefont {Chow},\ and\
  \citenamefont {Gambetta}}]{takita2016demonstration}%
  \BibitemOpen
  \bibfield  {author} {\bibinfo {author} {\bibfnamefont {M.}~\bibnamefont
  {Takita}}, \bibinfo {author} {\bibfnamefont {A.~D.}\ \bibnamefont
  {C{\'o}rcoles}}, \bibinfo {author} {\bibfnamefont {E.}~\bibnamefont
  {Magesan}}, \bibinfo {author} {\bibfnamefont {B.}~\bibnamefont {Abdo}},
  \bibinfo {author} {\bibfnamefont {M.}~\bibnamefont {Brink}}, \bibinfo
  {author} {\bibfnamefont {A.}~\bibnamefont {Cross}}, \bibinfo {author}
  {\bibfnamefont {J.~M.}\ \bibnamefont {Chow}}, \ and\ \bibinfo {author}
  {\bibfnamefont {J.~M.}\ \bibnamefont {Gambetta}},\ }\href@noop {} {\bibfield
  {journal} {\bibinfo  {journal} {Physical Review Letters}\ }\textbf {\bibinfo
  {volume} {117}},\ \bibinfo {pages} {210505} (\bibinfo {year}
  {2016})}\BibitemShut {NoStop}%
\bibitem [{\citenamefont {McKay}\ \emph {et~al.}(2017)\citenamefont {McKay},
  \citenamefont {Wood}, \citenamefont {Sheldon}, \citenamefont {Chow},\ and\
  \citenamefont {Gambetta}}]{mckay2017efficient}%
  \BibitemOpen
  \bibfield  {author} {\bibinfo {author} {\bibfnamefont {D.~C.}\ \bibnamefont
  {McKay}}, \bibinfo {author} {\bibfnamefont {C.~J.}\ \bibnamefont {Wood}},
  \bibinfo {author} {\bibfnamefont {S.}~\bibnamefont {Sheldon}}, \bibinfo
  {author} {\bibfnamefont {J.~M.}\ \bibnamefont {Chow}}, \ and\ \bibinfo
  {author} {\bibfnamefont {J.~M.}\ \bibnamefont {Gambetta}},\ }\href@noop {}
  {\bibfield  {journal} {\bibinfo  {journal} {Physical Review A}\ }\textbf
  {\bibinfo {volume} {96}},\ \bibinfo {pages} {022330} (\bibinfo {year}
  {2017})}\BibitemShut {NoStop}%
\bibitem [{\citenamefont {Bravyi}\ and\ \citenamefont
  {Kitaev}(1998)}]{bravyi1998quantum}%
  \BibitemOpen
  \bibfield  {author} {\bibinfo {author} {\bibfnamefont {S.~B.}\ \bibnamefont
  {Bravyi}}\ and\ \bibinfo {author} {\bibfnamefont {A.~Y.}\ \bibnamefont
  {Kitaev}},\ }\href@noop {} {\bibfield  {journal} {\bibinfo  {journal} {arXiv
  preprint quant-ph/9811052}\ } (\bibinfo {year} {1998})}\BibitemShut {NoStop}%
\bibitem [{\citenamefont {Fowler}\ \emph {et~al.}(2012)\citenamefont {Fowler},
  \citenamefont {Mariantoni}, \citenamefont {Martinis},\ and\ \citenamefont
  {Cleland}}]{fowler2012surface}%
  \BibitemOpen
  \bibfield  {author} {\bibinfo {author} {\bibfnamefont {A.~G.}\ \bibnamefont
  {Fowler}}, \bibinfo {author} {\bibfnamefont {M.}~\bibnamefont {Mariantoni}},
  \bibinfo {author} {\bibfnamefont {J.~M.}\ \bibnamefont {Martinis}}, \ and\
  \bibinfo {author} {\bibfnamefont {A.~N.}\ \bibnamefont {Cleland}},\
  }\href@noop {} {\bibfield  {journal} {\bibinfo  {journal} {Physical Review
  A}\ }\textbf {\bibinfo {volume} {86}},\ \bibinfo {pages} {032324} (\bibinfo
  {year} {2012})}\BibitemShut {NoStop}%
\bibitem [{\citenamefont {Werschnik}\ and\ \citenamefont
  {Gross}(2007)}]{werschnik2007quantum}%
  \BibitemOpen
  \bibfield  {author} {\bibinfo {author} {\bibfnamefont {J.}~\bibnamefont
  {Werschnik}}\ and\ \bibinfo {author} {\bibfnamefont {E.}~\bibnamefont
  {Gross}},\ }\href@noop {} {\bibfield  {journal} {\bibinfo  {journal} {Journal
  of Physics B: Atomic, Molecular and Optical Physics}\ }\textbf {\bibinfo
  {volume} {40}},\ \bibinfo {pages} {R175} (\bibinfo {year}
  {2007})}\BibitemShut {NoStop}%
\bibitem [{\citenamefont {Doria}\ \emph {et~al.}(2011)\citenamefont {Doria},
  \citenamefont {Calarco},\ and\ \citenamefont
  {Montangero}}]{PhysRevLett.106.190501}%
  \BibitemOpen
  \bibfield  {author} {\bibinfo {author} {\bibfnamefont {P.}~\bibnamefont
  {Doria}}, \bibinfo {author} {\bibfnamefont {T.}~\bibnamefont {Calarco}}, \
  and\ \bibinfo {author} {\bibfnamefont {S.}~\bibnamefont {Montangero}},\
  }\href {\doibase 10.1103/PhysRevLett.106.190501} {\bibfield  {journal}
  {\bibinfo  {journal} {Physical Review Letters}\ }\textbf {\bibinfo {volume}
  {106}},\ \bibinfo {pages} {190501} (\bibinfo {year} {2011})}\BibitemShut
  {NoStop}%
\bibitem [{\citenamefont {Rach}\ \emph {et~al.}(2015)\citenamefont {Rach},
  \citenamefont {M\"uller}, \citenamefont {Calarco},\ and\ \citenamefont
  {Montangero}}]{PhysRevA.92.062343}%
  \BibitemOpen
  \bibfield  {author} {\bibinfo {author} {\bibfnamefont {N.}~\bibnamefont
  {Rach}}, \bibinfo {author} {\bibfnamefont {M.~M.}\ \bibnamefont {M\"uller}},
  \bibinfo {author} {\bibfnamefont {T.}~\bibnamefont {Calarco}}, \ and\
  \bibinfo {author} {\bibfnamefont {S.}~\bibnamefont {Montangero}},\ }\href
  {\doibase 10.1103/PhysRevA.92.062343} {\bibfield  {journal} {\bibinfo
  {journal} {Physical Review A}\ }\textbf {\bibinfo {volume} {92}},\ \bibinfo
  {pages} {062343} (\bibinfo {year} {2015})}\BibitemShut {NoStop}%
\bibitem [{\citenamefont {Khaneja}\ \emph {et~al.}(2005)\citenamefont
  {Khaneja}, \citenamefont {Reiss}, \citenamefont {Kehlet}, \citenamefont
  {Schulte-Herbr{\"u}ggen},\ and\ \citenamefont {Glaser}}]{khaneja2005optimal}%
  \BibitemOpen
  \bibfield  {author} {\bibinfo {author} {\bibfnamefont {N.}~\bibnamefont
  {Khaneja}}, \bibinfo {author} {\bibfnamefont {T.}~\bibnamefont {Reiss}},
  \bibinfo {author} {\bibfnamefont {C.}~\bibnamefont {Kehlet}}, \bibinfo
  {author} {\bibfnamefont {T.}~\bibnamefont {Schulte-Herbr{\"u}ggen}}, \ and\
  \bibinfo {author} {\bibfnamefont {S.~J.}\ \bibnamefont {Glaser}},\
  }\href@noop {} {\bibfield  {journal} {\bibinfo  {journal} {Journal of
  Magnetic Resonance}\ }\textbf {\bibinfo {volume} {172}},\ \bibinfo {pages}
  {296} (\bibinfo {year} {2005})}\BibitemShut {NoStop}%
\bibitem [{\citenamefont {Li}\ \emph {et~al.}(2017)\citenamefont {Li},
  \citenamefont {Yang}, \citenamefont {Peng},\ and\ \citenamefont
  {Sun}}]{li2017hybrid}%
  \BibitemOpen
  \bibfield  {author} {\bibinfo {author} {\bibfnamefont {J.}~\bibnamefont
  {Li}}, \bibinfo {author} {\bibfnamefont {X.}~\bibnamefont {Yang}}, \bibinfo
  {author} {\bibfnamefont {X.}~\bibnamefont {Peng}}, \ and\ \bibinfo {author}
  {\bibfnamefont {C.-P.}\ \bibnamefont {Sun}},\ }\href@noop {} {\bibfield
  {journal} {\bibinfo  {journal} {Physical Review Letters}\ }\textbf {\bibinfo
  {volume} {118}},\ \bibinfo {pages} {150503} (\bibinfo {year}
  {2017})}\BibitemShut {NoStop}%
\bibitem [{\citenamefont {Konnov}\ and\ \citenamefont
  {Krotov}(1999)}]{konnov1999global}%
  \BibitemOpen
  \bibfield  {author} {\bibinfo {author} {\bibfnamefont {A.}~\bibnamefont
  {Konnov}}\ and\ \bibinfo {author} {\bibfnamefont {V.~F.}\ \bibnamefont
  {Krotov}},\ }\href@noop {} {\bibfield  {journal} {\bibinfo  {journal}
  {Avtomatika i Telemekhanika}\ }\textbf {\bibinfo {volume} {10}},\ \bibinfo
  {pages} {77} (\bibinfo {year} {1999})}\BibitemShut {NoStop}%
\bibitem [{\citenamefont {Sklarz}\ and\ \citenamefont
  {Tannor}(2002)}]{sklarz2002loading}%
  \BibitemOpen
  \bibfield  {author} {\bibinfo {author} {\bibfnamefont {S.~E.}\ \bibnamefont
  {Sklarz}}\ and\ \bibinfo {author} {\bibfnamefont {D.~J.}\ \bibnamefont
  {Tannor}},\ }\href@noop {} {\bibfield  {journal} {\bibinfo  {journal}
  {Physical Review A}\ }\textbf {\bibinfo {volume} {66}},\ \bibinfo {pages}
  {053619} (\bibinfo {year} {2002})}\BibitemShut {NoStop}%
\bibitem [{\citenamefont {Reich}\ \emph {et~al.}(2012)\citenamefont {Reich},
  \citenamefont {Ndong},\ and\ \citenamefont {Koch}}]{reich2012monotonically}%
  \BibitemOpen
  \bibfield  {author} {\bibinfo {author} {\bibfnamefont {D.~M.}\ \bibnamefont
  {Reich}}, \bibinfo {author} {\bibfnamefont {M.}~\bibnamefont {Ndong}}, \ and\
  \bibinfo {author} {\bibfnamefont {C.~P.}\ \bibnamefont {Koch}},\ }\href@noop
  {} {\bibfield  {journal} {\bibinfo  {journal} {The Journal of Chemical
  Physics}\ }\textbf {\bibinfo {volume} {136}},\ \bibinfo {pages} {104103}
  (\bibinfo {year} {2012})}\BibitemShut {NoStop}%
\bibitem [{\citenamefont {Machnes}\ \emph {et~al.}(2018)\citenamefont
  {Machnes}, \citenamefont {Ass{\'e}mat}, \citenamefont {Tannor},\ and\
  \citenamefont {Wilhelm}}]{machnes2018tunable}%
  \BibitemOpen
  \bibfield  {author} {\bibinfo {author} {\bibfnamefont {S.}~\bibnamefont
  {Machnes}}, \bibinfo {author} {\bibfnamefont {E.}~\bibnamefont
  {Ass{\'e}mat}}, \bibinfo {author} {\bibfnamefont {D.}~\bibnamefont {Tannor}},
  \ and\ \bibinfo {author} {\bibfnamefont {F.~K.}\ \bibnamefont {Wilhelm}},\
  }\href@noop {} {\bibfield  {journal} {\bibinfo  {journal} {Physical Review
  Letters}\ }\textbf {\bibinfo {volume} {120}},\ \bibinfo {pages} {150401}
  (\bibinfo {year} {2018})}\BibitemShut {NoStop}%
\bibitem [{\citenamefont {Kirchhoff}\ \emph {et~al.}(2018)\citenamefont
  {Kirchhoff}, \citenamefont {Ke{\ss}ler}, \citenamefont {Liebermann},
  \citenamefont {Ass{\'e}mat}, \citenamefont {Machnes}, \citenamefont
  {Motzoi},\ and\ \citenamefont {Wilhelm}}]{kirchhoff2018optimized}%
  \BibitemOpen
  \bibfield  {author} {\bibinfo {author} {\bibfnamefont {S.}~\bibnamefont
  {Kirchhoff}}, \bibinfo {author} {\bibfnamefont {T.}~\bibnamefont
  {Ke{\ss}ler}}, \bibinfo {author} {\bibfnamefont {P.~J.}\ \bibnamefont
  {Liebermann}}, \bibinfo {author} {\bibfnamefont {E.}~\bibnamefont
  {Ass{\'e}mat}}, \bibinfo {author} {\bibfnamefont {S.}~\bibnamefont
  {Machnes}}, \bibinfo {author} {\bibfnamefont {F.}~\bibnamefont {Motzoi}}, \
  and\ \bibinfo {author} {\bibfnamefont {F.~K.}\ \bibnamefont {Wilhelm}},\
  }\href@noop {} {\bibfield  {journal} {\bibinfo  {journal} {Physical Review
  A}\ }\textbf {\bibinfo {volume} {97}},\ \bibinfo {pages} {042348} (\bibinfo
  {year} {2018})}\BibitemShut {NoStop}%
\bibitem [{\citenamefont {S{\o}rensen}\ \emph {et~al.}(2018)\citenamefont
  {S{\o}rensen}, \citenamefont {Aranburu}, \citenamefont {Heinzel},\ and\
  \citenamefont {Sherson}}]{sorensen2018gradient}%
  \BibitemOpen
  \bibfield  {author} {\bibinfo {author} {\bibfnamefont {J.~J.}\ \bibnamefont
  {S{\o}rensen}}, \bibinfo {author} {\bibfnamefont {M.}~\bibnamefont
  {Aranburu}}, \bibinfo {author} {\bibfnamefont {T.}~\bibnamefont {Heinzel}}, \
  and\ \bibinfo {author} {\bibfnamefont {J.}~\bibnamefont {Sherson}},\
  }\href@noop {} {\bibfield  {journal} {\bibinfo  {journal} {arXiv preprint
  arXiv:1802.07509}\ } (\bibinfo {year} {2018})}\BibitemShut {NoStop}%
\bibitem [{\citenamefont {Nocedal}\ and\ \citenamefont
  {Wright}(2006)}]{Nocedal2006NO}%
  \BibitemOpen
  \bibfield  {author} {\bibinfo {author} {\bibfnamefont {J.}~\bibnamefont
  {Nocedal}}\ and\ \bibinfo {author} {\bibfnamefont {S.~J.}\ \bibnamefont
  {Wright}},\ }\href@noop {} {\emph {\bibinfo {title} {Numerical
  Optimization}}},\ \bibinfo {edition} {\emph{2nd}}\ ed.\ (\bibinfo
  {publisher} {Springer},\ \bibinfo {address} {New York},\ \bibinfo {year}
  {2006})\BibitemShut {NoStop}%
\bibitem [{\citenamefont {Tannor}\ and\ \citenamefont
  {Rice}(1985)}]{tannor1985control}%
  \BibitemOpen
  \bibfield  {author} {\bibinfo {author} {\bibfnamefont {D.~J.}\ \bibnamefont
  {Tannor}}\ and\ \bibinfo {author} {\bibfnamefont {S.~A.}\ \bibnamefont
  {Rice}},\ }\href@noop {} {\bibfield  {journal} {\bibinfo  {journal} {The
  Journal of Chemical Physics}\ }\textbf {\bibinfo {volume} {83}},\ \bibinfo
  {pages} {5013} (\bibinfo {year} {1985})}\BibitemShut {NoStop}%
\bibitem [{\citenamefont {Conolly}\ \emph {et~al.}(1986)\citenamefont
  {Conolly}, \citenamefont {Nishimura},\ and\ \citenamefont
  {Macovski}}]{conolly1986optimal}%
  \BibitemOpen
  \bibfield  {author} {\bibinfo {author} {\bibfnamefont {S.}~\bibnamefont
  {Conolly}}, \bibinfo {author} {\bibfnamefont {D.}~\bibnamefont {Nishimura}},
  \ and\ \bibinfo {author} {\bibfnamefont {A.}~\bibnamefont {Macovski}},\
  }\href@noop {} {\bibfield  {journal} {\bibinfo  {journal} {IEEE Transactions
  on Medical Imaging}\ }\textbf {\bibinfo {volume} {5}},\ \bibinfo {pages}
  {106} (\bibinfo {year} {1986})}\BibitemShut {NoStop}%
\bibitem [{\citenamefont {Peirce}\ \emph {et~al.}(1988)\citenamefont {Peirce},
  \citenamefont {Dahleh},\ and\ \citenamefont {Rabitz}}]{peirce1988optimal}%
  \BibitemOpen
  \bibfield  {author} {\bibinfo {author} {\bibfnamefont {A.~P.}\ \bibnamefont
  {Peirce}}, \bibinfo {author} {\bibfnamefont {M.~A.}\ \bibnamefont {Dahleh}},
  \ and\ \bibinfo {author} {\bibfnamefont {H.}~\bibnamefont {Rabitz}},\
  }\href@noop {} {\bibfield  {journal} {\bibinfo  {journal} {Physical Review
  A}\ }\textbf {\bibinfo {volume} {37}},\ \bibinfo {pages} {4950} (\bibinfo
  {year} {1988})}\BibitemShut {NoStop}%
\bibitem [{\citenamefont {Zahedinejad}\ \emph {et~al.}(2015)\citenamefont
  {Zahedinejad}, \citenamefont {Ghosh},\ and\ \citenamefont
  {Sanders}}]{zahedinejad2015high}%
  \BibitemOpen
  \bibfield  {author} {\bibinfo {author} {\bibfnamefont {E.}~\bibnamefont
  {Zahedinejad}}, \bibinfo {author} {\bibfnamefont {J.}~\bibnamefont {Ghosh}},
  \ and\ \bibinfo {author} {\bibfnamefont {B.~C.}\ \bibnamefont {Sanders}},\
  }\href@noop {} {\bibfield  {journal} {\bibinfo  {journal} {Physical Review
  Letters}\ }\textbf {\bibinfo {volume} {114}},\ \bibinfo {pages} {200502}
  (\bibinfo {year} {2015})}\BibitemShut {NoStop}%
\bibitem [{\citenamefont {Wu}\ \emph {et~al.}(2018)\citenamefont {Wu},
  \citenamefont {Chu}, \citenamefont {Owens},\ and\ \citenamefont
  {Rabitz}}]{wu2018data}%
  \BibitemOpen
  \bibfield  {author} {\bibinfo {author} {\bibfnamefont {R.-B.}\ \bibnamefont
  {Wu}}, \bibinfo {author} {\bibfnamefont {B.}~\bibnamefont {Chu}}, \bibinfo
  {author} {\bibfnamefont {D.~H.}\ \bibnamefont {Owens}}, \ and\ \bibinfo
  {author} {\bibfnamefont {H.}~\bibnamefont {Rabitz}},\ }\href@noop {}
  {\bibfield  {journal} {\bibinfo  {journal} {Physical Review A}\ }\textbf
  {\bibinfo {volume} {97}},\ \bibinfo {pages} {042122} (\bibinfo {year}
  {2018})}\BibitemShut {NoStop}%
\bibitem [{\citenamefont {Kosloff}\ \emph {et~al.}(1989)\citenamefont
  {Kosloff}, \citenamefont {Rice}, \citenamefont {Gaspard}, \citenamefont
  {Tersigni},\ and\ \citenamefont {Tannor}}]{kosloff1989wavepacket}%
  \BibitemOpen
  \bibfield  {author} {\bibinfo {author} {\bibfnamefont {R.}~\bibnamefont
  {Kosloff}}, \bibinfo {author} {\bibfnamefont {S.~A.}\ \bibnamefont {Rice}},
  \bibinfo {author} {\bibfnamefont {P.}~\bibnamefont {Gaspard}}, \bibinfo
  {author} {\bibfnamefont {S.}~\bibnamefont {Tersigni}}, \ and\ \bibinfo
  {author} {\bibfnamefont {D.}~\bibnamefont {Tannor}},\ }\href@noop {}
  {\bibfield  {journal} {\bibinfo  {journal} {Chemical Physics}\ }\textbf
  {\bibinfo {volume} {139}},\ \bibinfo {pages} {201} (\bibinfo {year}
  {1989})}\BibitemShut {NoStop}%
\bibitem [{\citenamefont {Brixner}\ \emph {et~al.}(2001)\citenamefont
  {Brixner}, \citenamefont {Damrauer}, \citenamefont {Niklaus},\ and\
  \citenamefont {Gerber}}]{brixner2001photoselective}%
  \BibitemOpen
  \bibfield  {author} {\bibinfo {author} {\bibfnamefont {T.}~\bibnamefont
  {Brixner}}, \bibinfo {author} {\bibfnamefont {N.}~\bibnamefont {Damrauer}},
  \bibinfo {author} {\bibfnamefont {P.}~\bibnamefont {Niklaus}}, \ and\
  \bibinfo {author} {\bibfnamefont {G.}~\bibnamefont {Gerber}},\ }\href@noop {}
  {\bibfield  {journal} {\bibinfo  {journal} {Nature}\ }\textbf {\bibinfo
  {volume} {414}},\ \bibinfo {pages} {57} (\bibinfo {year} {2001})}\BibitemShut
  {NoStop}%
\bibitem [{\citenamefont {Bartels}\ \emph {et~al.}(2000)\citenamefont
  {Bartels}, \citenamefont {Backus}, \citenamefont {Zeek}, \citenamefont
  {Misoguti}, \citenamefont {Vdovin}, \citenamefont {Christov}, \citenamefont
  {Murnane},\ and\ \citenamefont {Kapteyn}}]{bartels2000shaped}%
  \BibitemOpen
  \bibfield  {author} {\bibinfo {author} {\bibfnamefont {R.}~\bibnamefont
  {Bartels}}, \bibinfo {author} {\bibfnamefont {S.}~\bibnamefont {Backus}},
  \bibinfo {author} {\bibfnamefont {E.}~\bibnamefont {Zeek}}, \bibinfo {author}
  {\bibfnamefont {L.}~\bibnamefont {Misoguti}}, \bibinfo {author}
  {\bibfnamefont {G.}~\bibnamefont {Vdovin}}, \bibinfo {author} {\bibfnamefont
  {I.}~\bibnamefont {Christov}}, \bibinfo {author} {\bibfnamefont
  {M.}~\bibnamefont {Murnane}}, \ and\ \bibinfo {author} {\bibfnamefont
  {H.}~\bibnamefont {Kapteyn}},\ }\href@noop {} {\bibfield  {journal} {\bibinfo
   {journal} {Nature}\ }\textbf {\bibinfo {volume} {406}},\ \bibinfo {pages}
  {164} (\bibinfo {year} {2000})}\BibitemShut {NoStop}%
\bibitem [{\citenamefont {Buckup}\ \emph {et~al.}(2006)\citenamefont {Buckup},
  \citenamefont {Lebold}, \citenamefont {Weigel}, \citenamefont {Wohlleben},\
  and\ \citenamefont {Motzkus}}]{buckup2006singlet}%
  \BibitemOpen
  \bibfield  {author} {\bibinfo {author} {\bibfnamefont {T.}~\bibnamefont
  {Buckup}}, \bibinfo {author} {\bibfnamefont {T.}~\bibnamefont {Lebold}},
  \bibinfo {author} {\bibfnamefont {A.}~\bibnamefont {Weigel}}, \bibinfo
  {author} {\bibfnamefont {W.}~\bibnamefont {Wohlleben}}, \ and\ \bibinfo
  {author} {\bibfnamefont {M.}~\bibnamefont {Motzkus}},\ }\href@noop {}
  {\bibfield  {journal} {\bibinfo  {journal} {Journal of Photochemistry and
  Photobiology A: Chemistry}\ }\textbf {\bibinfo {volume} {180}},\ \bibinfo
  {pages} {314} (\bibinfo {year} {2006})}\BibitemShut {NoStop}%
\bibitem [{\citenamefont {Haessler}\ \emph {et~al.}(2014)\citenamefont
  {Haessler}, \citenamefont {Bal{\v{c}}iunas}, \citenamefont {Fan},
  \citenamefont {Andriukaitis}, \citenamefont {Pug{\v{z}}lys}, \citenamefont
  {Baltu{\v{s}}ka}, \citenamefont {Witting}, \citenamefont {Squibb},
  \citenamefont {Za{\"\i}r}, \citenamefont {Tisch}, \citenamefont {Marangos},\
  and\ \citenamefont {Baltu{\v{s}}ka}}]{haessler2014optimization}%
  \BibitemOpen
  \bibfield  {author} {\bibinfo {author} {\bibfnamefont {S.}~\bibnamefont
  {Haessler}}, \bibinfo {author} {\bibfnamefont {T.}~\bibnamefont
  {Bal{\v{c}}iunas}}, \bibinfo {author} {\bibfnamefont {G.}~\bibnamefont
  {Fan}}, \bibinfo {author} {\bibfnamefont {G.}~\bibnamefont {Andriukaitis}},
  \bibinfo {author} {\bibfnamefont {A.}~\bibnamefont {Pug{\v{z}}lys}}, \bibinfo
  {author} {\bibfnamefont {A.}~\bibnamefont {Baltu{\v{s}}ka}}, \bibinfo
  {author} {\bibfnamefont {T.}~\bibnamefont {Witting}}, \bibinfo {author}
  {\bibfnamefont {R.}~\bibnamefont {Squibb}}, \bibinfo {author} {\bibfnamefont
  {A.}~\bibnamefont {Za{\"\i}r}}, \bibinfo {author} {\bibfnamefont {J.~W.}\
  \bibnamefont {Tisch}}, \bibinfo {author} {\bibfnamefont {J.~P.}\ \bibnamefont
  {Marangos}}, \ and\ \bibinfo {author} {\bibfnamefont {A.}~\bibnamefont
  {Baltu{\v{s}}ka}},\ }\href@noop {} {\bibfield  {journal} {\bibinfo  {journal}
  {Physical Review X}\ }\textbf {\bibinfo {volume} {4}},\ \bibinfo {pages}
  {021028} (\bibinfo {year} {2014})}\BibitemShut {NoStop}%
\bibitem [{\citenamefont {Dolde}\ \emph {et~al.}(2014)\citenamefont {Dolde},
  \citenamefont {Bergholm}, \citenamefont {Wang}, \citenamefont {Jakobi},
  \citenamefont {Naydenov}, \citenamefont {Pezzagna}, \citenamefont {Meijer},
  \citenamefont {Jelezko}, \citenamefont {Neumann}, \citenamefont
  {Schulte-Herbr{\"u}ggen}, \citenamefont {Biamonte},\ and\ \citenamefont
  {Wrachtrup}}]{dolde2014high}%
  \BibitemOpen
  \bibfield  {author} {\bibinfo {author} {\bibfnamefont {F.}~\bibnamefont
  {Dolde}}, \bibinfo {author} {\bibfnamefont {V.}~\bibnamefont {Bergholm}},
  \bibinfo {author} {\bibfnamefont {Y.}~\bibnamefont {Wang}}, \bibinfo {author}
  {\bibfnamefont {I.}~\bibnamefont {Jakobi}}, \bibinfo {author} {\bibfnamefont
  {B.}~\bibnamefont {Naydenov}}, \bibinfo {author} {\bibfnamefont
  {S.}~\bibnamefont {Pezzagna}}, \bibinfo {author} {\bibfnamefont
  {J.}~\bibnamefont {Meijer}}, \bibinfo {author} {\bibfnamefont
  {F.}~\bibnamefont {Jelezko}}, \bibinfo {author} {\bibfnamefont
  {P.}~\bibnamefont {Neumann}}, \bibinfo {author} {\bibfnamefont
  {T.}~\bibnamefont {Schulte-Herbr{\"u}ggen}}, \bibinfo {author} {\bibfnamefont
  {J.}~\bibnamefont {Biamonte}}, \ and\ \bibinfo {author} {\bibfnamefont
  {J.}~\bibnamefont {Wrachtrup}},\ }\href@noop {} {\bibfield  {journal}
  {\bibinfo  {journal} {Nature Communications}\ }\textbf {\bibinfo {volume}
  {5}},\ \bibinfo {pages} {3371} (\bibinfo {year} {2014})}\BibitemShut
  {NoStop}%
\bibitem [{\citenamefont {Heeres}\ \emph {et~al.}(2017)\citenamefont {Heeres},
  \citenamefont {Reinhold}, \citenamefont {Ofek}, \citenamefont {Frunzio},
  \citenamefont {Jiang}, \citenamefont {Devoret},\ and\ \citenamefont
  {Schoelkopf}}]{heeres2017implementing}%
  \BibitemOpen
  \bibfield  {author} {\bibinfo {author} {\bibfnamefont {R.~W.}\ \bibnamefont
  {Heeres}}, \bibinfo {author} {\bibfnamefont {P.}~\bibnamefont {Reinhold}},
  \bibinfo {author} {\bibfnamefont {N.}~\bibnamefont {Ofek}}, \bibinfo {author}
  {\bibfnamefont {L.}~\bibnamefont {Frunzio}}, \bibinfo {author} {\bibfnamefont
  {L.}~\bibnamefont {Jiang}}, \bibinfo {author} {\bibfnamefont {M.~H.}\
  \bibnamefont {Devoret}}, \ and\ \bibinfo {author} {\bibfnamefont {R.~J.}\
  \bibnamefont {Schoelkopf}},\ }\href@noop {} {\bibfield  {journal} {\bibinfo
  {journal} {Nature Communications}\ }\textbf {\bibinfo {volume} {8}},\
  \bibinfo {pages} {94} (\bibinfo {year} {2017})}\BibitemShut {NoStop}%
\bibitem [{\citenamefont {Feng}\ \emph {et~al.}(2018)\citenamefont {Feng},
  \citenamefont {Cho}, \citenamefont {Katiyar}, \citenamefont {Li},
  \citenamefont {Lu}, \citenamefont {Baugh},\ and\ \citenamefont
  {Laflamme}}]{feng2018closed}%
  \BibitemOpen
  \bibfield  {author} {\bibinfo {author} {\bibfnamefont {G.}~\bibnamefont
  {Feng}}, \bibinfo {author} {\bibfnamefont {F.~H.}\ \bibnamefont {Cho}},
  \bibinfo {author} {\bibfnamefont {H.}~\bibnamefont {Katiyar}}, \bibinfo
  {author} {\bibfnamefont {J.}~\bibnamefont {Li}}, \bibinfo {author}
  {\bibfnamefont {D.}~\bibnamefont {Lu}}, \bibinfo {author} {\bibfnamefont
  {J.}~\bibnamefont {Baugh}}, \ and\ \bibinfo {author} {\bibfnamefont
  {R.}~\bibnamefont {Laflamme}},\ }\href@noop {} {\bibfield  {journal}
  {\bibinfo  {journal} {arXiv preprint arXiv:1805.11674}\ } (\bibinfo {year}
  {2018})}\BibitemShut {NoStop}%
\bibitem [{\citenamefont {Lu}\ \emph {et~al.}(2017)\citenamefont {Lu},
  \citenamefont {Li}, \citenamefont {Li}, \citenamefont {Katiyar},
  \citenamefont {Park}, \citenamefont {Feng}, \citenamefont {Xin},
  \citenamefont {Li}, \citenamefont {Long}, \citenamefont {Brodutch},
  \citenamefont {Baugh}, \citenamefont {Zeng},\ and\ \citenamefont
  {Laflamme}}]{lu2017enhancing}%
  \BibitemOpen
  \bibfield  {author} {\bibinfo {author} {\bibfnamefont {D.}~\bibnamefont
  {Lu}}, \bibinfo {author} {\bibfnamefont {K.}~\bibnamefont {Li}}, \bibinfo
  {author} {\bibfnamefont {J.}~\bibnamefont {Li}}, \bibinfo {author}
  {\bibfnamefont {H.}~\bibnamefont {Katiyar}}, \bibinfo {author} {\bibfnamefont
  {A.~J.}\ \bibnamefont {Park}}, \bibinfo {author} {\bibfnamefont
  {G.}~\bibnamefont {Feng}}, \bibinfo {author} {\bibfnamefont {T.}~\bibnamefont
  {Xin}}, \bibinfo {author} {\bibfnamefont {H.}~\bibnamefont {Li}}, \bibinfo
  {author} {\bibfnamefont {G.}~\bibnamefont {Long}}, \bibinfo {author}
  {\bibfnamefont {A.}~\bibnamefont {Brodutch}}, \bibinfo {author}
  {\bibfnamefont {J.}~\bibnamefont {Baugh}}, \bibinfo {author} {\bibfnamefont
  {B.}~\bibnamefont {Zeng}}, \ and\ \bibinfo {author} {\bibfnamefont
  {R.}~\bibnamefont {Laflamme}},\ }\href@noop {} {\bibfield  {journal}
  {\bibinfo  {journal} {npj Quantum Information}\ }\textbf {\bibinfo {volume}
  {3}},\ \bibinfo {pages} {45} (\bibinfo {year} {2017})}\BibitemShut {NoStop}%
\bibitem [{\citenamefont {Rigetti}\ and\ \citenamefont
  {Devoret}(2010)}]{rigetti2010fully}%
  \BibitemOpen
  \bibfield  {author} {\bibinfo {author} {\bibfnamefont {C.}~\bibnamefont
  {Rigetti}}\ and\ \bibinfo {author} {\bibfnamefont {M.}~\bibnamefont
  {Devoret}},\ }\href@noop {} {\bibfield  {journal} {\bibinfo  {journal}
  {Physical Review B}\ }\textbf {\bibinfo {volume} {81}},\ \bibinfo {pages}
  {134507} (\bibinfo {year} {2010})}\BibitemShut {NoStop}%
\bibitem [{\citenamefont {Nielsen}\ and\ \citenamefont
  {Chuang}(2002)}]{nielsen2002quantum}%
  \BibitemOpen
  \bibfield  {author} {\bibinfo {author} {\bibfnamefont {M.~A.}\ \bibnamefont
  {Nielsen}}\ and\ \bibinfo {author} {\bibfnamefont {I.}~\bibnamefont
  {Chuang}},\ }\href@noop {} {\emph {\bibinfo {title} {Quantum Computation and
  Quantum Information}}}\ (\bibinfo  {publisher} {AAPT},\ \bibinfo {year}
  {2002})\BibitemShut {NoStop}%
\bibitem [{\citenamefont {Flammia}\ and\ \citenamefont
  {Liu}(2011)}]{PhysRevLett.106.230501}%
  \BibitemOpen
  \bibfield  {author} {\bibinfo {author} {\bibfnamefont {S.~T.}\ \bibnamefont
  {Flammia}}\ and\ \bibinfo {author} {\bibfnamefont {Y.-K.}\ \bibnamefont
  {Liu}},\ }\href {\doibase 10.1103/PhysRevLett.106.230501} {\bibfield
  {journal} {\bibinfo  {journal} {Physical Review Letters}\ }\textbf {\bibinfo
  {volume} {106}},\ \bibinfo {pages} {230501} (\bibinfo {year}
  {2011})}\BibitemShut {NoStop}%
\bibitem [{\citenamefont {Heya}\ \emph {et~al.}()\citenamefont {Heya},
  \citenamefont {Suzuki}, \citenamefont {Nakamura},\ and\ \citenamefont
  {Fujii}}]{heya2018supli}%
  \BibitemOpen
  \bibfield  {author} {\bibinfo {author} {\bibfnamefont {K.}~\bibnamefont
  {Heya}}, \bibinfo {author} {\bibfnamefont {Y.}~\bibnamefont {Suzuki}},
  \bibinfo {author} {\bibfnamefont {Y.}~\bibnamefont {Nakamura}}, \ and\
  \bibinfo {author} {\bibfnamefont {K.}~\bibnamefont {Fujii}},\ }\href@noop {}
  {\enquote {\bibinfo {title} {supplementary information},}\ }\BibitemShut
  {NoStop}%
\bibitem [{\citenamefont {Zanardi}(2001)}]{zanardi2001entanglement}%
  \BibitemOpen
  \bibfield  {author} {\bibinfo {author} {\bibfnamefont {P.}~\bibnamefont
  {Zanardi}},\ }\href@noop {} {\bibfield  {journal} {\bibinfo  {journal}
  {Physical Review A}\ }\textbf {\bibinfo {volume} {63}},\ \bibinfo {pages}
  {040304} (\bibinfo {year} {2001})}\BibitemShut {NoStop}%
\bibitem [{\citenamefont {Sousa}\ and\ \citenamefont
  {Ramos}(2006)}]{sousa2006universal}%
  \BibitemOpen
  \bibfield  {author} {\bibinfo {author} {\bibfnamefont {P.~B.}\ \bibnamefont
  {Sousa}}\ and\ \bibinfo {author} {\bibfnamefont {R.~V.}\ \bibnamefont
  {Ramos}},\ }\href@noop {} {\bibfield  {journal} {\bibinfo  {journal} {arXiv
  preprint quant-ph/0602174}\ } (\bibinfo {year} {2006})}\BibitemShut {NoStop}%
\bibitem [{\citenamefont {Nelder}\ and\ \citenamefont
  {Mead}(1965)}]{nelder1965simplex}%
  \BibitemOpen
  \bibfield  {author} {\bibinfo {author} {\bibfnamefont {J.~A.}\ \bibnamefont
  {Nelder}}\ and\ \bibinfo {author} {\bibfnamefont {R.}~\bibnamefont {Mead}},\
  }\href@noop {} {\bibfield  {journal} {\bibinfo  {journal} {The Computer
  Journal}\ }\textbf {\bibinfo {volume} {7}},\ \bibinfo {pages} {308} (\bibinfo
  {year} {1965})}\BibitemShut {NoStop}%
\bibitem [{\citenamefont {Powell}(1964)}]{powell1964efficient}%
  \BibitemOpen
  \bibfield  {author} {\bibinfo {author} {\bibfnamefont {M.~J.}\ \bibnamefont
  {Powell}},\ }\href@noop {} {\bibfield  {journal} {\bibinfo  {journal} {The
  Computer Journal}\ }\textbf {\bibinfo {volume} {7}},\ \bibinfo {pages} {155}
  (\bibinfo {year} {1964})}\BibitemShut {NoStop}%
\bibitem [{\citenamefont {Conn}\ \emph {et~al.}(1997)\citenamefont {Conn},
  \citenamefont {Scheinberg},\ and\ \citenamefont
  {Toint}}]{conn1997convergence}%
  \BibitemOpen
  \bibfield  {author} {\bibinfo {author} {\bibfnamefont {A.~R.}\ \bibnamefont
  {Conn}}, \bibinfo {author} {\bibfnamefont {K.}~\bibnamefont {Scheinberg}}, \
  and\ \bibinfo {author} {\bibfnamefont {P.~L.}\ \bibnamefont {Toint}},\
  }\href@noop {} {\emph {\bibinfo {title} {On the Convergence of
  Derivative-Free Methods for Unconstrained Optimization}}}\ (\bibinfo
  {publisher} {Cambridge University Press},\ \bibinfo {year} {1997})\ pp.\
  \bibinfo {pages} {83--108}\BibitemShut {NoStop}%
\bibitem [{\citenamefont {Fletcher}(2013)}]{fletcher2013practical}%
  \BibitemOpen
  \bibfield  {author} {\bibinfo {author} {\bibfnamefont {R.}~\bibnamefont
  {Fletcher}},\ }\href@noop {} {\emph {\bibinfo {title} {Practical Methods of
  Optimization}}}\ (\bibinfo  {publisher} {John Wiley \& Sons},\ \bibinfo
  {year} {2013})\BibitemShut {NoStop}%
\bibitem [{\citenamefont {Byrd}\ \emph {et~al.}(1995)\citenamefont {Byrd},
  \citenamefont {Lu}, \citenamefont {Nocedal},\ and\ \citenamefont
  {Zhu}}]{byrd1995limited}%
  \BibitemOpen
  \bibfield  {author} {\bibinfo {author} {\bibfnamefont {R.~H.}\ \bibnamefont
  {Byrd}}, \bibinfo {author} {\bibfnamefont {P.}~\bibnamefont {Lu}}, \bibinfo
  {author} {\bibfnamefont {J.}~\bibnamefont {Nocedal}}, \ and\ \bibinfo
  {author} {\bibfnamefont {C.}~\bibnamefont {Zhu}},\ }\href@noop {} {\bibfield
  {journal} {\bibinfo  {journal} {SIAM Journal on Scientific Computing}\
  }\textbf {\bibinfo {volume} {16}},\ \bibinfo {pages} {1190} (\bibinfo {year}
  {1995})}\BibitemShut {NoStop}%
\bibitem [{\citenamefont {Hestenes}\ and\ \citenamefont
  {Stiefel}(1952)}]{hestenes1952methods}%
  \BibitemOpen
  \bibfield  {author} {\bibinfo {author} {\bibfnamefont {M.~R.}\ \bibnamefont
  {Hestenes}}\ and\ \bibinfo {author} {\bibfnamefont {E.}~\bibnamefont
  {Stiefel}},\ }\href@noop {} {\emph {\bibinfo {title} {Methods of Conjugate
  Gradients for Solving Linear Systems}}},\ Vol.~\bibinfo {volume} {49}\
  (\bibinfo  {publisher} {NBS Washington, DC},\ \bibinfo {year}
  {1952})\BibitemShut {NoStop}%
\bibitem [{\citenamefont {Kingma}\ and\ \citenamefont
  {Ba}(2014)}]{kingma2014adam}%
  \BibitemOpen
  \bibfield  {author} {\bibinfo {author} {\bibfnamefont {D.~P.}\ \bibnamefont
  {Kingma}}\ and\ \bibinfo {author} {\bibfnamefont {J.}~\bibnamefont {Ba}},\
  }\href@noop {} {\bibfield  {journal} {\bibinfo  {journal} {arXiv preprint
  arXiv:1412.6980}\ } (\bibinfo {year} {2014})}\BibitemShut {NoStop}%
\bibitem [{\citenamefont {Spall}(1992)}]{119632}%
  \BibitemOpen
  \bibfield  {author} {\bibinfo {author} {\bibfnamefont {J.~C.}\ \bibnamefont
  {Spall}},\ }\href {\doibase 10.1109/9.119632} {\bibfield  {journal} {\bibinfo
   {journal} {IEEE Transactions on Automatic Control}\ }\textbf {\bibinfo
  {volume} {37}},\ \bibinfo {pages} {332} (\bibinfo {year} {1992})}\BibitemShut
  {NoStop}%
\bibitem [{\citenamefont {Russell}\ \emph {et~al.}(2018)\citenamefont
  {Russell}, \citenamefont {Vuglar},\ and\ \citenamefont
  {Rabitz}}]{russell2018control}%
  \BibitemOpen
  \bibfield  {author} {\bibinfo {author} {\bibfnamefont {B.~J.}\ \bibnamefont
  {Russell}}, \bibinfo {author} {\bibfnamefont {S.}~\bibnamefont {Vuglar}}, \
  and\ \bibinfo {author} {\bibfnamefont {H.}~\bibnamefont {Rabitz}},\
  }\href@noop {} {\bibfield  {journal} {\bibinfo  {journal} {Journal of Physics
  A: Mathematical and Theoretical}\ }\textbf {\bibinfo {volume} {51}} (\bibinfo
  {year} {2018})}\BibitemShut {NoStop}%
\bibitem [{\citenamefont {Chow}\ \emph {et~al.}(2011)\citenamefont {Chow},
  \citenamefont {C{\'o}rcoles}, \citenamefont {Gambetta}, \citenamefont
  {Rigetti}, \citenamefont {Johnson}, \citenamefont {Smolin}, \citenamefont
  {Rozen}, \citenamefont {Keefe}, \citenamefont {Rothwell}, \citenamefont
  {Ketchen},\ and\ \citenamefont {Steffen}}]{chow2011simple}%
  \BibitemOpen
  \bibfield  {author} {\bibinfo {author} {\bibfnamefont {J.~M.}\ \bibnamefont
  {Chow}}, \bibinfo {author} {\bibfnamefont {A.}~\bibnamefont {C{\'o}rcoles}},
  \bibinfo {author} {\bibfnamefont {J.~M.}\ \bibnamefont {Gambetta}}, \bibinfo
  {author} {\bibfnamefont {C.}~\bibnamefont {Rigetti}}, \bibinfo {author}
  {\bibfnamefont {B.}~\bibnamefont {Johnson}}, \bibinfo {author} {\bibfnamefont
  {J.~A.}\ \bibnamefont {Smolin}}, \bibinfo {author} {\bibfnamefont
  {J.}~\bibnamefont {Rozen}}, \bibinfo {author} {\bibfnamefont {G.~A.}\
  \bibnamefont {Keefe}}, \bibinfo {author} {\bibfnamefont {M.~B.}\ \bibnamefont
  {Rothwell}}, \bibinfo {author} {\bibfnamefont {M.~B.}\ \bibnamefont
  {Ketchen}}, \ and\ \bibinfo {author} {\bibfnamefont {M.}~\bibnamefont
  {Steffen}},\ }\href@noop {} {\bibfield  {journal} {\bibinfo  {journal}
  {Physical Review Letters}\ }\textbf {\bibinfo {volume} {107}},\ \bibinfo
  {pages} {080502} (\bibinfo {year} {2011})}\BibitemShut {NoStop}%
\bibitem [{\citenamefont {Sheldon}\ \emph {et~al.}(2016)\citenamefont
  {Sheldon}, \citenamefont {Magesan}, \citenamefont {Chow},\ and\ \citenamefont
  {Gambetta}}]{sheldon2016procedure}%
  \BibitemOpen
  \bibfield  {author} {\bibinfo {author} {\bibfnamefont {S.}~\bibnamefont
  {Sheldon}}, \bibinfo {author} {\bibfnamefont {E.}~\bibnamefont {Magesan}},
  \bibinfo {author} {\bibfnamefont {J.~M.}\ \bibnamefont {Chow}}, \ and\
  \bibinfo {author} {\bibfnamefont {J.~M.}\ \bibnamefont {Gambetta}},\
  }\href@noop {} {\bibfield  {journal} {\bibinfo  {journal} {Physical Review
  A}\ }\textbf {\bibinfo {volume} {93}},\ \bibinfo {pages} {060302} (\bibinfo
  {year} {2016})}\BibitemShut {NoStop}%
\bibitem [{\citenamefont {Gottesman}(1997)}]{gottesman1997stabilizer}%
  \BibitemOpen
  \bibfield  {author} {\bibinfo {author} {\bibfnamefont {D.}~\bibnamefont
  {Gottesman}},\ }\href@noop {} {\bibfield  {journal} {\bibinfo  {journal}
  {arXiv preprint quant-ph/9705052}\ } (\bibinfo {year} {1997})}\BibitemShut
  {NoStop}%
\bibitem [{\citenamefont {Temme}\ \emph {et~al.}(2017)\citenamefont {Temme},
  \citenamefont {Bravyi},\ and\ \citenamefont {Gambetta}}]{temme2017error}%
  \BibitemOpen
  \bibfield  {author} {\bibinfo {author} {\bibfnamefont {K.}~\bibnamefont
  {Temme}}, \bibinfo {author} {\bibfnamefont {S.}~\bibnamefont {Bravyi}}, \
  and\ \bibinfo {author} {\bibfnamefont {J.~M.}\ \bibnamefont {Gambetta}},\
  }\href@noop {} {\bibfield  {journal} {\bibinfo  {journal} {Physical Review
  Letters}\ }\textbf {\bibinfo {volume} {119}},\ \bibinfo {pages} {180509}
  (\bibinfo {year} {2017})}\BibitemShut {NoStop}%
\bibitem [{\citenamefont {Endo}\ \emph {et~al.}(2018)\citenamefont {Endo},
  \citenamefont {Benjamin},\ and\ \citenamefont {Li}}]{endo2018practical}%
  \BibitemOpen
  \bibfield  {author} {\bibinfo {author} {\bibfnamefont {S.}~\bibnamefont
  {Endo}}, \bibinfo {author} {\bibfnamefont {S.~C.}\ \bibnamefont {Benjamin}},
  \ and\ \bibinfo {author} {\bibfnamefont {Y.}~\bibnamefont {Li}},\ }\href@noop
  {} {\bibfield  {journal} {\bibinfo  {journal} {Physical Review X}\ }\textbf
  {\bibinfo {volume} {8}},\ \bibinfo {pages} {031027} (\bibinfo {year}
  {2018})}\BibitemShut {NoStop}%
\bibitem [{\citenamefont {Li}\ and\ \citenamefont
  {Benjamin}(2017)}]{li2017efficient}%
  \BibitemOpen
  \bibfield  {author} {\bibinfo {author} {\bibfnamefont {Y.}~\bibnamefont
  {Li}}\ and\ \bibinfo {author} {\bibfnamefont {S.~C.}\ \bibnamefont
  {Benjamin}},\ }\href@noop {} {\bibfield  {journal} {\bibinfo  {journal}
  {Physical Review X}\ }\textbf {\bibinfo {volume} {7}},\ \bibinfo {pages}
  {021050} (\bibinfo {year} {2017})}\BibitemShut {NoStop}%
\bibitem [{\citenamefont {Merkel}\ \emph {et~al.}(2013)\citenamefont {Merkel},
  \citenamefont {Gambetta}, \citenamefont {Smolin}, \citenamefont {Poletto},
  \citenamefont {C{\'o}rcoles}, \citenamefont {Johnson}, \citenamefont {Ryan},\
  and\ \citenamefont {Steffen}}]{merkel2013self}%
  \BibitemOpen
  \bibfield  {author} {\bibinfo {author} {\bibfnamefont {S.~T.}\ \bibnamefont
  {Merkel}}, \bibinfo {author} {\bibfnamefont {J.~M.}\ \bibnamefont
  {Gambetta}}, \bibinfo {author} {\bibfnamefont {J.~A.}\ \bibnamefont
  {Smolin}}, \bibinfo {author} {\bibfnamefont {S.}~\bibnamefont {Poletto}},
  \bibinfo {author} {\bibfnamefont {A.~D.}\ \bibnamefont {C{\'o}rcoles}},
  \bibinfo {author} {\bibfnamefont {B.~R.}\ \bibnamefont {Johnson}}, \bibinfo
  {author} {\bibfnamefont {C.~A.}\ \bibnamefont {Ryan}}, \ and\ \bibinfo
  {author} {\bibfnamefont {M.}~\bibnamefont {Steffen}},\ }\href@noop {}
  {\bibfield  {journal} {\bibinfo  {journal} {Physical Review A}\ }\textbf
  {\bibinfo {volume} {87}},\ \bibinfo {pages} {062119} (\bibinfo {year}
  {2013})}\BibitemShut {NoStop}%
\bibitem [{\citenamefont {Blume-Kohout}\ \emph {et~al.}(2013)\citenamefont
  {Blume-Kohout}, \citenamefont {Gamble}, \citenamefont {Nielsen},
  \citenamefont {Mizrahi}, \citenamefont {Sterk},\ and\ \citenamefont
  {Maunz}}]{blume2013robust}%
  \BibitemOpen
  \bibfield  {author} {\bibinfo {author} {\bibfnamefont {R.}~\bibnamefont
  {Blume-Kohout}}, \bibinfo {author} {\bibfnamefont {J.~K.}\ \bibnamefont
  {Gamble}}, \bibinfo {author} {\bibfnamefont {E.}~\bibnamefont {Nielsen}},
  \bibinfo {author} {\bibfnamefont {J.}~\bibnamefont {Mizrahi}}, \bibinfo
  {author} {\bibfnamefont {J.~D.}\ \bibnamefont {Sterk}}, \ and\ \bibinfo
  {author} {\bibfnamefont {P.}~\bibnamefont {Maunz}},\ }\href@noop {}
  {\bibfield  {journal} {\bibinfo  {journal} {arXiv preprint arXiv:1310.4492}\
  } (\bibinfo {year} {2013})}\BibitemShut {NoStop}%
\bibitem [{\citenamefont {Sugiyama}\ \emph {et~al.}(2018)\citenamefont
  {Sugiyama}, \citenamefont {Imori},\ and\ \citenamefont
  {Tanaka}}]{sugiyama2018reliable}%
  \BibitemOpen
  \bibfield  {author} {\bibinfo {author} {\bibfnamefont {T.}~\bibnamefont
  {Sugiyama}}, \bibinfo {author} {\bibfnamefont {S.}~\bibnamefont {Imori}}, \
  and\ \bibinfo {author} {\bibfnamefont {F.}~\bibnamefont {Tanaka}},\
  }\href@noop {} {\bibfield  {journal} {\bibinfo  {journal} {arXiv preprint
  arXiv:1806.02696}\ } (\bibinfo {year} {2018})}\BibitemShut {NoStop}%
\bibitem [{\citenamefont {Johansson}\ \emph {et~al.}(2012)\citenamefont
  {Johansson}, \citenamefont {Nation},\ and\ \citenamefont
  {Nori}}]{johansson2012qutip}%
  \BibitemOpen
  \bibfield  {author} {\bibinfo {author} {\bibfnamefont {J.}~\bibnamefont
  {Johansson}}, \bibinfo {author} {\bibfnamefont {P.}~\bibnamefont {Nation}}, \
  and\ \bibinfo {author} {\bibfnamefont {F.}~\bibnamefont {Nori}},\ }\href@noop
  {} {\bibfield  {journal} {\bibinfo  {journal} {Computer Physics
  Communications}\ }\textbf {\bibinfo {volume} {183}},\ \bibinfo {pages} {1760}
  (\bibinfo {year} {2012})}\BibitemShut {NoStop}%
\bibitem [{\citenamefont {Jones}\ \emph {et~al.}(01  )\citenamefont {Jones},
  \citenamefont {Oliphant},\ and\ \citenamefont {Peterson}}]{scipy}%
  \BibitemOpen
  \bibfield  {author} {\bibinfo {author} {\bibfnamefont {E.}~\bibnamefont
  {Jones}}, \bibinfo {author} {\bibfnamefont {T.}~\bibnamefont {Oliphant}}, \
  and\ \bibinfo {author} {\bibfnamefont {P.}~\bibnamefont {Peterson}},\ }\href
  {"http://www.scipy.org/"} {\enquote {\bibinfo {title} {{SciPy}: Open source
  scientific tools for {Python}},}\ } (\bibinfo {year} {2001--})\BibitemShut
  {NoStop}%
\bibitem [{\citenamefont {Abadi}\ \emph {et~al.}(2015)\citenamefont {Abadi},
  \citenamefont {Agarwal}, \citenamefont {Barham}, \citenamefont {Brevdo},
  \citenamefont {Chen}, \citenamefont {Citro}, \citenamefont {Corrado},
  \citenamefont {Davis}, \citenamefont {Dean}, \citenamefont {Devin},
  \citenamefont {Ghemawat}, \citenamefont {Goodfellow}, \citenamefont {Harp},
  \citenamefont {Irving}, \citenamefont {Isard}, \citenamefont {Jia},
  \citenamefont {Jozefowicz}, \citenamefont {Kaiser}, \citenamefont {Kudlur},
  \citenamefont {Levenberg}, \citenamefont {Man\'{e}}, \citenamefont {Monga},
  \citenamefont {Moore}, \citenamefont {Murray}, \citenamefont {Olah},
  \citenamefont {Schuster}, \citenamefont {Shlens}, \citenamefont {Steiner},
  \citenamefont {Sutskever}, \citenamefont {Talwar}, \citenamefont {Tucker},
  \citenamefont {Vanhoucke}, \citenamefont {Vasudevan}, \citenamefont
  {Vi\'{e}gas}, \citenamefont {Vinyals}, \citenamefont {Warden}, \citenamefont
  {Wattenberg}, \citenamefont {Wicke}, \citenamefont {Yu},\ and\ \citenamefont
  {Zheng}}]{tensorflow2015-whitepaper}%
  \BibitemOpen
  \bibfield  {author} {\bibinfo {author} {\bibfnamefont {M.}~\bibnamefont
  {Abadi}}, \bibinfo {author} {\bibfnamefont {A.}~\bibnamefont {Agarwal}},
  \bibinfo {author} {\bibfnamefont {P.}~\bibnamefont {Barham}}, \bibinfo
  {author} {\bibfnamefont {E.}~\bibnamefont {Brevdo}}, \bibinfo {author}
  {\bibfnamefont {Z.}~\bibnamefont {Chen}}, \bibinfo {author} {\bibfnamefont
  {C.}~\bibnamefont {Citro}}, \bibinfo {author} {\bibfnamefont {G.~S.}\
  \bibnamefont {Corrado}}, \bibinfo {author} {\bibfnamefont {A.}~\bibnamefont
  {Davis}}, \bibinfo {author} {\bibfnamefont {J.}~\bibnamefont {Dean}},
  \bibinfo {author} {\bibfnamefont {M.}~\bibnamefont {Devin}}, \bibinfo
  {author} {\bibfnamefont {S.}~\bibnamefont {Ghemawat}}, \bibinfo {author}
  {\bibfnamefont {I.}~\bibnamefont {Goodfellow}}, \bibinfo {author}
  {\bibfnamefont {A.}~\bibnamefont {Harp}}, \bibinfo {author} {\bibfnamefont
  {G.}~\bibnamefont {Irving}}, \bibinfo {author} {\bibfnamefont
  {M.}~\bibnamefont {Isard}}, \bibinfo {author} {\bibfnamefont
  {Y.}~\bibnamefont {Jia}}, \bibinfo {author} {\bibfnamefont {R.}~\bibnamefont
  {Jozefowicz}}, \bibinfo {author} {\bibfnamefont {L.}~\bibnamefont {Kaiser}},
  \bibinfo {author} {\bibfnamefont {M.}~\bibnamefont {Kudlur}}, \bibinfo
  {author} {\bibfnamefont {J.}~\bibnamefont {Levenberg}}, \bibinfo {author}
  {\bibfnamefont {D.}~\bibnamefont {Man\'{e}}}, \bibinfo {author}
  {\bibfnamefont {R.}~\bibnamefont {Monga}}, \bibinfo {author} {\bibfnamefont
  {S.}~\bibnamefont {Moore}}, \bibinfo {author} {\bibfnamefont
  {D.}~\bibnamefont {Murray}}, \bibinfo {author} {\bibfnamefont
  {C.}~\bibnamefont {Olah}}, \bibinfo {author} {\bibfnamefont {M.}~\bibnamefont
  {Schuster}}, \bibinfo {author} {\bibfnamefont {J.}~\bibnamefont {Shlens}},
  \bibinfo {author} {\bibfnamefont {B.}~\bibnamefont {Steiner}}, \bibinfo
  {author} {\bibfnamefont {I.}~\bibnamefont {Sutskever}}, \bibinfo {author}
  {\bibfnamefont {K.}~\bibnamefont {Talwar}}, \bibinfo {author} {\bibfnamefont
  {P.}~\bibnamefont {Tucker}}, \bibinfo {author} {\bibfnamefont
  {V.}~\bibnamefont {Vanhoucke}}, \bibinfo {author} {\bibfnamefont
  {V.}~\bibnamefont {Vasudevan}}, \bibinfo {author} {\bibfnamefont
  {F.}~\bibnamefont {Vi\'{e}gas}}, \bibinfo {author} {\bibfnamefont
  {O.}~\bibnamefont {Vinyals}}, \bibinfo {author} {\bibfnamefont
  {P.}~\bibnamefont {Warden}}, \bibinfo {author} {\bibfnamefont
  {M.}~\bibnamefont {Wattenberg}}, \bibinfo {author} {\bibfnamefont
  {M.}~\bibnamefont {Wicke}}, \bibinfo {author} {\bibfnamefont
  {Y.}~\bibnamefont {Yu}}, \ and\ \bibinfo {author} {\bibfnamefont
  {X.}~\bibnamefont {Zheng}},\ }\href {https://www.tensorflow.org/} {\enquote
  {\bibinfo {title} {{TensorFlow}: Large-scale machine learning on
  heterogeneous systems},}\ } (\bibinfo {year} {2015}),\ \bibinfo {note}
  {software available from tensorflow.org}\BibitemShut {NoStop}%
\bibitem [{\citenamefont {Helgason}(2001)}]{helgason2001differential}%
  \BibitemOpen
  \bibfield  {author} {\bibinfo {author} {\bibfnamefont {S.}~\bibnamefont
  {Helgason}},\ }\href@noop {} {\emph {\bibinfo {title} {Differential Geometry
  and Symmetric Spaces}}},\ Vol.\ \bibinfo {volume} {341}\ (\bibinfo
  {publisher} {American Mathematical Soc.},\ \bibinfo {year}
  {2001})\BibitemShut {NoStop}%
\bibitem [{\citenamefont {Nielsen}\ \emph {et~al.}(2003)\citenamefont
  {Nielsen}, \citenamefont {Dawson}, \citenamefont {Dodd}, \citenamefont
  {Gilchrist}, \citenamefont {Mortimer}, \citenamefont {Osborne}, \citenamefont
  {Bremner}, \citenamefont {Harrow},\ and\ \citenamefont
  {Hines}}]{nielsen2003quantum}%
  \BibitemOpen
  \bibfield  {author} {\bibinfo {author} {\bibfnamefont {M.~A.}\ \bibnamefont
  {Nielsen}}, \bibinfo {author} {\bibfnamefont {C.~M.}\ \bibnamefont {Dawson}},
  \bibinfo {author} {\bibfnamefont {J.~L.}\ \bibnamefont {Dodd}}, \bibinfo
  {author} {\bibfnamefont {A.}~\bibnamefont {Gilchrist}}, \bibinfo {author}
  {\bibfnamefont {D.}~\bibnamefont {Mortimer}}, \bibinfo {author}
  {\bibfnamefont {T.~J.}\ \bibnamefont {Osborne}}, \bibinfo {author}
  {\bibfnamefont {M.~J.}\ \bibnamefont {Bremner}}, \bibinfo {author}
  {\bibfnamefont {A.~W.}\ \bibnamefont {Harrow}}, \ and\ \bibinfo {author}
  {\bibfnamefont {A.}~\bibnamefont {Hines}},\ }\href@noop {} {\bibfield
  {journal} {\bibinfo  {journal} {Physical Review A}\ }\textbf {\bibinfo
  {volume} {67}},\ \bibinfo {pages} {052301} (\bibinfo {year}
  {2003})}\BibitemShut {NoStop}%
\end{thebibliography}%
	
\end{document}